\documentclass{aa}

\usepackage{graphicx}

\usepackage{txfonts}
\usepackage{xcolor}

\newcommand{\xqr}{E-XQR-30}

\newcommand{\civ}{C\,{\sc iv}}
\newcommand{\siiv}{Si\,{\sc iv}}
\newcommand{\cf}{$\xi(v, v+\Delta v)$}
\newcommand{\md}{$\Omega_{\rm C\,\textsc{iv}}$}

\newcommand{\kms}{km\,s$^{-1}$}

\begin{document}

   \title{The clustering of \civ\ and \siiv\ at the end of reionisation}

   \subtitle{A perspective from the E-XQR-30 survey}

   \author{Louise Welsh\inst{1, 2, 3} \fnmsep\thanks{Email: louise.welsh@inaf.it}
          \and
          Valentina D'Odorico\inst{1,2,4}
          \and
          Fabio Fontanot\inst{1,2}
          \and 
          Rebecca Davies\inst{5}
          \and 
          Sarah E. I. Bosman\inst{6,7}
          \and
           Guido Cupani\inst{1,2}
          \and
          George Becker\inst{8}
          \and 
          Laura Keating\inst{9}
          \and 
          Emma Ryan-Weber\inst{5,10}
          \and 
          Manuela Bischetti\inst{11}
          \and 
          Martin Haehnelt\inst{12,13}
          \and
          Huanqing Chen\inst{14} 
          \and
          Yongda Zhu\inst{15}
          \and
          Samuel Lai\inst{16}
          \and 
          Michaela Hirschmann\inst{1}
          \and 
          Lizhi Xie\inst{17}
          \and 
          Yuxiang Qin\inst{18}
          }

\institute{\inst{1} INAF - Osservatorio Astronomico di Trieste, via G. B. Tiepolo 11, I-34143 Trieste, Italy \\
\inst{2} IFPU - Institute for Fundamental Physics of the Universe, via Beirut 2, I-34151 Trieste, Italy \\
\inst{3} Centre for Extragalactic Astronomy, Durham University, South Road, Durham DH1 3LE, UK \\
\inst{4} Scuola Normale Superiore, Piazza dei Cavalieri 7, I-56126 Pisa, Italy \\
\inst{5}  Centre for Astrophysics and Supercomputing, Swinburne University of Technology, Hawthorn, Victoria 3122, Australia \\
\inst{6} Institute for Theoretical Physics, Heidelberg University, Philosophenweg 12, D–69120, Heidelberg, Germany \\
\inst{7} Max-Planck-Institut f\"{u}r Astronomie, K\"{o}nigstuhl 17, 69117 Heidelberg, Germany \\
\inst{8} Department of Physics \& Astronomy, University of California, Riverside, CA 92521, USA \\
\inst{9} Institute for Astronomy, University of Edinburgh, Blackford Hill, Edinburgh, EH9 3HJ, UK \\
\inst{10} ARC Centre of Excellence for All Sky Astrophysics in 3 Dimensions (ASTRO3D), Canberra, ACT 2611, Australia \\
\inst{11} Dipartimento di Fisica, Universitá di Trieste, Sezione di Astronomia, Via G.B. Tiepolo 11, I-34131 Trieste, Italy \\
\inst{12} Institute of Astronomy, University of Cambridge, Madingley Road, Cambridge CB3 0HA, UK \\
\inst{13} Kavli Institute for Cosmology, University of Cambridge, Madingley Road, Cambridge CB3 0HA, UK \\
\inst{14} Augustana Campus, University of Alberta, Camrose, AB T4V2R3, Canada \\
\inst{15} Steward Observatory, University of Arizona, 933 North Cherry Avenue, Tucson, AZ 85721, USA \\
\inst{16} Commonwealth Scientific and Industrial Research Organisation (CSIRO), Space \& Astronomy, P. O. Box 1130, Bentley, WA 6102, Australia \\
\inst{17} Tianjin Normal University, Binshuixidao 393, 300387, Tianjin, China \\
\inst{18} Research School of Astronomy and Astrophysics, Australian National University, Canberra, ACT 2611, Australia 
}

   \date{Received ---; accepted ---}

  \abstract
{} 
{We aim to study the clustering of metal absorption lines and the structures that they arise in as a function of cosmic time. We focus on the behaviour of \civ\ and \siiv\ ionic species. These \civ\ and \siiv\ absorption features are identified along a given quasar sightline.}
{We exploit the two-point correlation function (2PCF) to investigate the clustering of these structures as a function of their separation. We utilise the E-XQR-30 data to perform a novel analysis at $z>5$. We also draw on literature surveys (including XQ-100) of lower redshift quasars to investigate the possible evolution of this clustering towards cosmic noon (i.e., $z\sim 2-3$). }
{We find no significant evolution with redshift when considering the separation of absorbers in velocity space. Since we are comparing data across a large interval of cosmic time, we also consider the separation between absorbers in the reference frame of physical distances. 
In this reference frame, we find that the amplitude of the clustering increases with cosmic time for both \civ\ and \siiv\ on the scales of $<1500$~ physical kpc. }
{For the first time, we assess the 2PCF of \civ\ and \siiv\ close to the epoch of reionisation utilising the absorber catalogue from the E-XQR-30 survey. We compare this with lower redshift data and find that, on small scales, the clustering of these structures grows with cosmic time. We compare these results to the clustering of galaxies in the GAEA simulations. It appears that the structures traced by \civ\ 
are broadly comparable to the galaxies from the considered simulations. The clustering is most similar to that of the galaxies with virial masses $M \sim10^{10.5}~{\rm M}_{\odot}$. 
We do not draw direct comparisons at the smallest separations to avoid the clustering traced by \civ\ at $z\sim5$ being dominated by contributions from absorbers within a single halo. We require tailor-made simulations to investigate the full range of factors contributing to the observed clustering of the detected metal absorbers. Future ground-based spectrographs will further facilitate surveys of absorbers at this epoch with increased sensitivity.}


   \maketitle
%

\section{Introduction}
Quasar absorption line spectroscopy is one of the most powerful tools currently at our disposal to observe the gas in and around galaxies.
The circumgalactic medium (CGM) constitutes a large fraction of the baryonic mass in dark matter haloes and is intrinsically linked with galaxy evolution \citep{Tumlinson2017}, while the intergalactic medium (IGM) allows us to trace the evolution of gas and dust between galaxies \citep{Peroux2020}. The observed metals provide information on the properties of stars and galaxies in the early Universe. 
Crucially, the metals that are detected are often associated with structures that have no detectable counterpart in emission. Thus, this technique allows us to study some of the lowest density structures in the early Universe.

The triply ionised carbon doublet (\civ\ $\lambda \lambda$ 1548, 1551 \AA) is the most commonly detected metal absorption feature at $z>1$ \citep[e.g.][]{Lu1991, Cowie1995}. In particular, it can be used to trace the metallicity of the IGM as well as the overdense regions in the vicinity of galaxies. Indeed, there have been many efforts to simulate the carbon footprint of galactic halos \citep{Bird2016, Finlator2020}. 

While less common, \siiv\ is another invaluable tracer of ionised gas. The \siiv\ $\lambda \lambda$ 1394, 1403 \AA\ doublet shares some similar characteristics, that are useful for detection, to that of \civ. In particular, they are both found redward of the Ly$\alpha$ 1215~\AA\ line and can be detected independently due to their doublet nature. This means that the redshift of these absorption features can be determined in the absence of other associated metal transitions. Given the lower ionisation energy of \siiv, it has the potential to trace denser gas than that of \civ. Though, \civ\ and \siiv\ are often found concurrently in gas with similar densities and temperatures. This is especially true in comparison to other commonly detected metal transitions like O\,{\sc i} and Mg\,{\sc ii}.

In recent years, the discovery of quasars close the epoch of reionisation has become more efficient \citep[e.g.,][]{Banados2016, Yang2023}. This has led to a rapid increase in number of metal absorbers discovered at this epoch.  One such survey that has been extremely successful in providing a high-resolution, high signal-to-noise  sample of metal absorbers at this epoch is the E-XQR-30 survey \citep{Dodorico2023}. This sample combines an ESO Large Programme (PI: D'Odorico) dedicated to observing 30 of the brightest quasars known above $z>5.8$ with  VLT-XSHOOTER \citep{Vernet2011} with 12 archival spectra of similar quality. 

A fundamental measure of chemical evolution is the mass density of metals \citep[e.g., \md; ][]{Songaila2001, Pettini2003, Boksenberg2003, Dodorico2010, Dodorico2013}. Studies focused on $z>5$ have recently shown that the mass density of higher ionisation states like \civ\ and \siiv\ declines as the epoch of reionisation approaches \citep{Dodorico2022, Christensen2023, Davies2023b}. At the same time, the mass density of low-ionisation species (e.g., O\,{\sc i} and C\,{\sc ii}) appears to increase \citep{Becker2019, Sebastian2024}. This is in line with pioneering works which relied on fewer statistics \citep[e.g.,][]{Simcoe2006, RyanWeber2006, RyanWeber2009, Becker2009, Bosman2017}.  This may suggest that the \civ\ being traced is predominantly produced by metals that formed within a galaxy. However, the ability of galaxies to retain gas is highly dependent on both the stellar mass and feedback efficiency.

Another important statistic is the clustering of these metals. The clustering of \civ\ and \siiv\ metal absorption features can be investigated through the two-point correlation function (2PCF hereafter) \cf. This is a measure of how these metals are distributed in relation to one another. Deviations from a random distribution can reveal the structure of the cosmic web. Indeed, \civ\ has also be used to trace the size of the CGM through the association with nearby Lyman break galaxies \citep{Adelberger2005, Steidel2010} and the transverse autocorrelation function using close quasar pairs \citep[e.g.,][]{Mintz2022} and lensed quasars \citep[e.g., ][]{Lopez2024}.
At $z\gtrsim 5$, the detection of associated galaxies becomes challenging. Nevertheless, some Lyman-$\alpha$ emitters and [O\,{\sc iii}] emitters have been found in the vicinity of \civ\ absorption \citep{Diaz2011, Diaz2021, Kashino2023}.

Comparisons with cosmological hydrodynamical simulations have shown that the distribution of \civ\ absorbers can be impacted by the nature of the feedback mechanisms \citep[e.g., AGN-driven outflows, galactic winds, stellar feedback; ][]{Tescari2011}. As discussed in \citet{Diaz2011}, the detection of \civ\ $\sim80$~physical kpc from a galaxy at $z = 5.7$ would suggest that galaxy-wide outflows were active at the end of the epoch of reionisation. Recent work indicates that ionised outflows may be stronger at the end of reionisation compared to later epochs \citep{Bischetti2022, Fiore2023, Tortosa2024}. Reproducing the observations of metals at $z>5$ is a challenging process; however, there are simulations tailored towards understanding the hosts and environments of metals at these early epochs \citep[e.g.,][]{Doughty2023}.

The clustering properties of intergalactic metals were studied in the redshift interval $1.7<z<4.5$ by \citet{Scannapieco2006} and later by \citet{Dodorico2010} using samples of absorbers with a large overlap. 
\citet{Scannapieco2006} found that the clustering properties were independent of the column density of the lines used. The analysis by \citet{Dodorico2010} found that \civ\ absorbers of higher column density were more strongly clustered on small scales relative to their lower column density counterparts (both defined below). This distinction can be explained through the way the samples are treated during each analysis. The result by \citet{Scannapieco2006}  considers either a cut in the minimum column density or a cut in the maximum column density. The resulting analysis indicates that the correlation function is not affected by the completeness or the presence of saturated lines. However, the analysis in \citet{Dodorico2010} creates two mutually exclusive subsamples based on the column density of \civ.  Weak absorbers are defined as those with $12 < \log_{10}\,N({\rm C \textsc{iv})/ cm^{-2} }< 13$ while strong absorbers are those with $13< \log_{10}\,N({\rm C \textsc{iv})/ cm^{-2} }< 15$. This approach is better suited to detecting any column density driven evolution in the 2PCF. Overall, this comparison highlights the delicate nature of this calculation. To draw meaningful comparisons, one must consider the sensitivity and completeness of the utilised data. In addition, it suggests that there may be a difference between the dominant source of the weak and strong \civ\ lines. 

In this paper, we present a novel analysis of the clustering of \civ\ and \siiv\ absorption line systems identified at the end of the epoch of reionisation. This analysis uses data from the E-XQR-30 survey and is based on a sample of \civ\ and \siiv\ absorbers with a median redshift of $\tilde{z}_{\rm abs} = 5.1 $ and $\tilde{z}_{\rm abs} = 5.5$ respectively. The E-XQR-30 data provide the largest uniform sample of $z>5.7$ quasars observed with sufficient resolution required to calculate the 2PCF of metal absorbers at this epoch. 
The total \civ\ path length covered by the \xqr\ survey data is $\Delta X = 280.3$. This quantity gives the total redshift search interval covered by the survey decoupled from cosmic expansion.  Using lower redshift data, we investigate the possible redshift evolution of the 2PCF. We further use this analysis to understand the structures giving rise to the observed metals.

We adopt a Planck cosmology throughout, with $H_{0} = 67.4\pm 0.5~{\rm km\,s^{-1}Mpc^{-1}}$, $\Omega_{\rm m} = 0.315 \pm 0.007$ and $h = 0.677$ \citep{Planck2018}. This paper is organised as follows: In Section~\ref{sec:data} we summarise the data used in this analysis. In Section~\ref{sec:model} we describe our analysis approach before presenting the results in Section~\ref{sec:res}. We present a discussion in Section~\ref{sec:disc} and draw conclusions in Section~\ref{sec:conc}.

\section{Data}
\label{sec:data} 
The principal sample of \civ\ absorbers used in this work are from the \xqr\ catalogue presented in \citet{Davies2023}. We consider only the absorbers in the `primary' sample and, following the convention of \citet{Davies2023b}, define a proximity region size of 10,000~\kms. This dataset is optimal to study the 2PCF of \civ\  and \siiv\ towards the end of the epoch of reionisation. In total there are 479 \civ\ absorbers that span the redshift interval $4.3 < z_{\rm abs} < 6.8$. This study indicates that the data are 50 per cent complete for absorbers with $\log_{10}\,N({\rm C \textsc{iv})/ cm^{-2} }>13.22$. When investigating the clustering of structures through absorption features, it is essential that we are sensitive to the majority of features that we are characterising. We therefore impose a 50 per cent completeness criteria and consider only the 2PCF of the strong absorbers with $\log_{10}\,N({\rm C \textsc{iv})/ cm^{-2} }>13.22$. This ensures we are not biasing the resulting correlation function due to non-detections of the weakest lines. This is the most conservative completeness criteria offered in \citet{Davies2023} for the full sample --- it is the completeness criteria based on the ‘high redshift’ subsample of the catalogue. We are therefore at least 50 per cent complete at all wavelengths. We note that, with the next generation of ground-based telescopes, it will be possible to increase the completeness of the metal line absorber catalogues compiled at $z>5$ \citep{Dodorico2024}.

\begin{figure}
    \centering
    \includegraphics[width=\columnwidth]{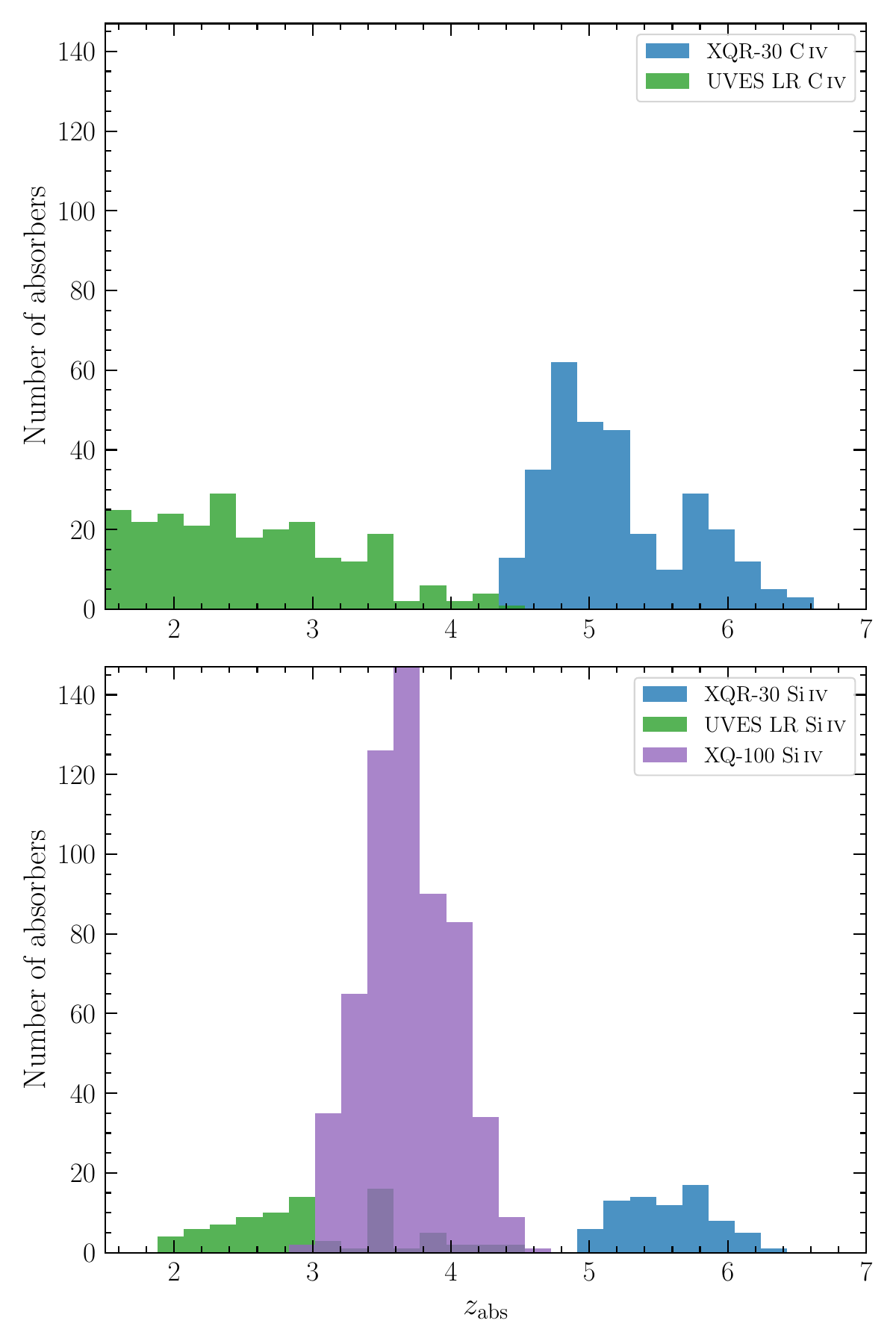}
    \caption{Redshift distribution of absorption features across the three absorber catalogues used in this work. The datasets are indicated in the legend. Note that there are distinctly more \civ\ absorbers than \siiv\ absorbers for a given dataset.}
    \label{fig:zabs}
\end{figure}

The \xqr\ metal absorption line catalogue also contains 127 \siiv\ absorbers in the interval $4.9 < z_{\rm abs} < 6.7$. Compared to \civ, the \siiv\ absorbers provide less information since the sample size is almost 4 times smaller. This limitation is particularly noticeable when considering pairs of absorbers that are found within small separations. However, the respective clustering of \civ\ and \siiv\  have  been found to trace each other closely \citep[e.g,][]{Scannapieco2006}. We therefore use \siiv\ as a useful benchmark of comparison. For these \siiv\ data, we adopt the 50 per cent completeness criteria $\log_{10}\,N({\rm Si \textsc{iv})/ cm^{-2} }>12.63$. There are ultimately 300 \civ\  and 76 \siiv\ usable features from this catalogue. Note that these usable features are necessarily found along a line-of-sight with at least two absorption features of the same transition that meet the completeness criteria. 
The entire sample of quasars from \xqr\ have usable \civ\ features. 21 of these quasars meet the same criteria for \siiv. 

We are primarily interested in comparing how the 2PCF of \civ\ may evolve across cosmic time. It is possible to split the E-XQR-30 data into multiple redshift bins. However, the most stringent statistics are provided when considering the full sample simultaneously. When performing these calculations for multiple redshift bins, we found no significant evolution over the redshift interval covered by XQR-30. This was limited by the large associated errors on the resulting 2PCFs. The median absorption redshift of the \civ\ features in the E-XQR-30 data is $\tilde{z}_{\rm abs}=5.1$ and for \siiv\ it is $\Tilde{z}_{\rm abs}=5.5$. 
To analyse the possible redshift evolution of the clustering of absorbers, we instead utilise two other datasets as points of comparison. In particular, we consider a compilation of \civ\ and \siiv\ features present in a sample of quasars observed with the Ultraviolet and Visual Echelle Spectrograph \citep[UVES; ][]{Dekker2000}. 
The details of these compilations are presented in \citet{Dodorico2010} and \citet{Dodorico2022}, respectively. The median absorption redshift of these features are $\tilde{z}_{\rm abs}=2.4$ and $\Tilde{z}_{\rm abs}=2.9$. We include a further sample of \siiv\ absorbers, also presented in \citet{Dodorico2022}, that are drawn from the XQ-100 survey \citep{Lopez2016}. The XQ-100 survey is a VLT-XSHOOTER survey of 100 quasars in the redshift interval $3.5 < z< 4.5$. 
The median redshift of the \siiv\ absorbers in XQ-100 is $\Tilde{z}_{\rm abs} = 3.7$. These data therefore produce insight into the behaviour of \siiv\ in the intermediate period between the end of the epoch or reionisation and cosmic noon. There is no equivalent catalogue of \civ\ absorbers from this dataset that could be used in this analysis. For all datasets, we impose the same cuts in column density as applied to the \xqr\ data.

When comparing these data, we must take into account any differences caused by the respective instruments used to observe these objects.  UVES data have a resolution of $R\sim 40\,000$ while XSHOOTER data have a resolution of $R\sim12\,000$. To account for this, we merge the lines present in the UVES data that would fall within one resolution element if observed with XSHOOTER.

We adopt a line merging procedure that has been used in previous works that compare similar datasets \citep[e.g.,][]{Dodorico2022}. We iteratively scan the catalogue for absorbers within $\Delta v = 50$~\kms\ of one another. These systems are merged into one absorber with a column density equivalent to the sum of the original lines and a redshift that is the column density weighted average of the lines. This process is repeated until there are no longer absorbers within $50$~\kms\ of one another. Note that the completeness criteria is applied after the line merging procedure.

Figure~\ref{fig:zabs} shows the distribution of absorption redshifts across the three datasets after applying the appropriate merging procedures. Here, and throughout, the sample drawn from merged UVES absorption lines is denoted as `UVES LR' given the high resolution of the original data.

\begin{figure}
	\includegraphics[width=\linewidth]{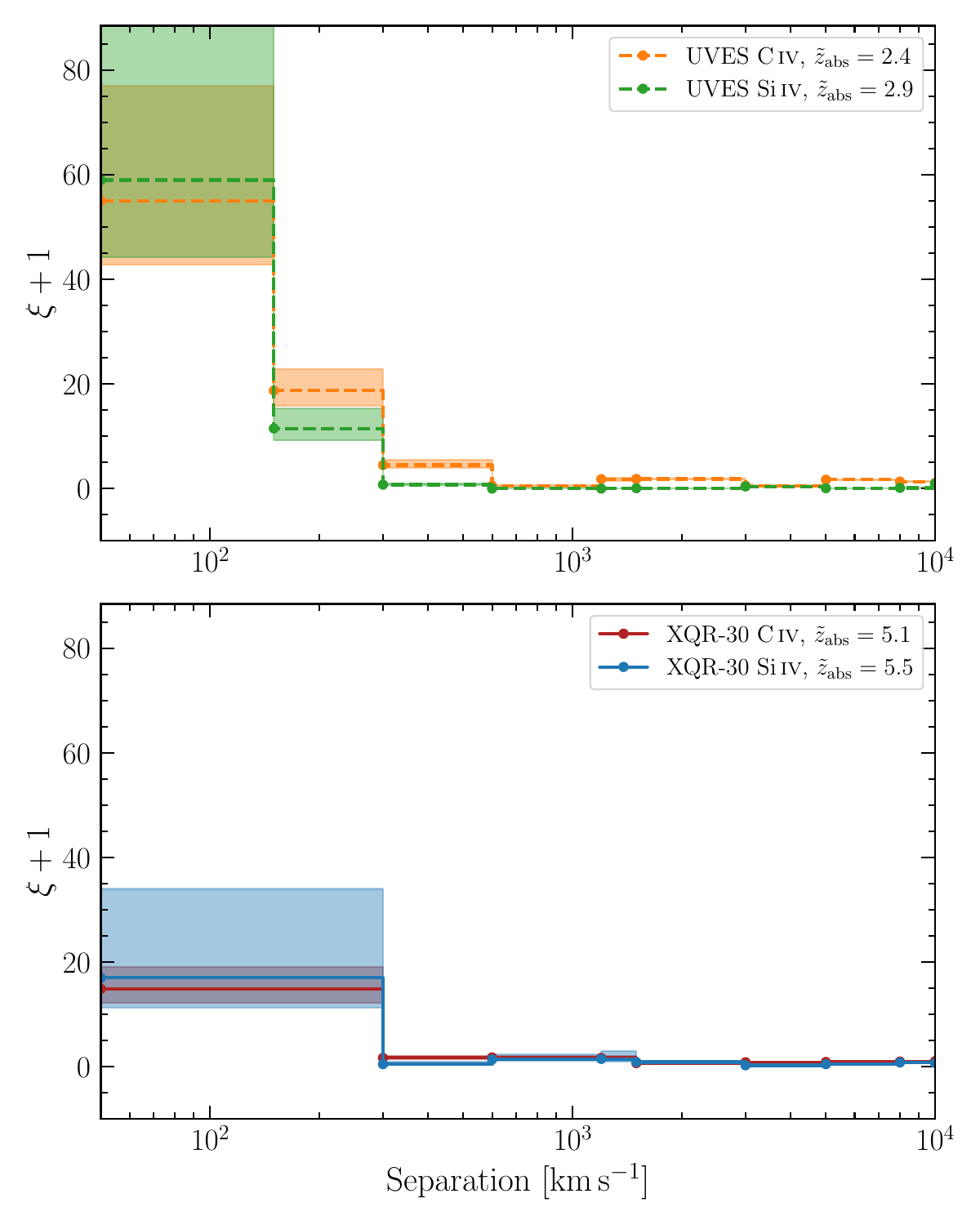}
    \caption{Top: The two point correlation function of \civ\ and \siiv\ based on the UVES LP data previously analysed in \citet{Dodorico2010, Dodorico2022}. In this  calculation we have imposed the same column density cuts as applied to the \xqr\ data. Compared to \xqr, the resulting distributions are sampled more frequently, with an additional  evaluation point at $\Delta v = 150$~\kms. The subsequent bins are consistent with those used for XQR-30. Bottom: The 2PCF of \civ\ and \siiv\ shown on the smallest scale accessible based on the limited statistics of the \siiv\ data. These bins span $\Delta v=$ [$50-300, 300-600, 600-1200, 1200-1500, 1500-3000$]~\kms\ for the first 5 bins. Note that distributions of $\xi$ values are formed via many realisations of the randomly distributed lines. For each sample, the solid line represents the median value while the shaded region encompasses the interquartile range.}
    \label{fig:2pcf_xqr30}
\end{figure}

\section{Analysis}
\label{sec:model}
The key quantity used to investigate the clustering of these metal absorbers is the two-point correlation function (recall 2PCF defined as $\xi$). To calculate this, we first compute the separation between all possible pairs of absorption features of a given species and ionisation state along a certain sightline. We then compare this to the separations that would be expected given a random distribution of absorbers. For each quasar, it is typical to calculate this in a sequence of velocity separation bins. Formally,
\begin{equation}
    \xi(v_{k}) + 1= \frac{n(v_{k})}{ \langle n_{\rm R}(v_{k}) \rangle }
\end{equation}
Where $v_{k} = v, v+\Delta v$. $n(v_{k})$ is the number of absorber pairs in the k-th bin of velocity separation. Subscript $R$ indicates the number of absorber pairs in the same velocity bin based on the random distribution of absorbers. Once summed over all quasars, this is:
\begin{equation}
\label{eqn:xi}
    \xi(v_{k})  + 1=  \frac{\sum\limits^{N_{\rm qso}}_{i=1} n(v_{k})_{i}}{\sum\limits^{N_{\rm qso}}_{i=1} \langle n_{\rm R}(v_{k})_{i} \rangle }
\end{equation}
In the above equation $N_{\rm qso}$ denotes number of individual sightlines. When looking at \civ\ from E-XQR-30, $N_{\rm qso}=42$. Note that we calculate the expected number of pairs in each separation bin for each quasar independently. First, we calculate the observed number density of lines across the quasar sample given the total redshift interval of the sample. For each quasar, we then calculate the expected number of lines given the redshift interval spanned by that sightline. The resulting separations are calculated after randomly populating the redshift interval with lines. We `throw' these lines sequentially following a uniform distribution in redshift space. If a line would be positioned within 50~\kms\ of a line that has already been generated, it is disregarded and thrown again. This is to ensure we are creating a random distribution in line with what is observable based on the real XSHOOTER data. To understand the uncertainty associated with these calculations, we generate 1000 random realisations of the fake absorption lines. Note that the numerator in Equation~\ref{eqn:xi} does not change during this process. For each realisation, we calculate the resulting correlation function. This builds a distribution of the correlation at each velocity separation that we consider. We then take the median value of this distribution to be representative of the correlation function and the errors are provided by the interquartile range of the distribution. We do not currently exclude potential broad absorption line (BAL) regions in the generated distributions of absorption lines. 

We have also tested a second approach for computing the distribution of feasible 2PCFs. We adopt a jackknife resampling technique to calculate the range of the 2PCFs given this sample of absorbers. The resulting distributions are indistinguishable from the nominal approach. This indicates that our analysis is not biased by an individual sightline in the sample.

\section{Results}
\label{sec:res}

Figure~\ref{fig:2pcf_xqr30} shows the resulting 2PCF based on the \xqr\ data and the data taken at lower redshift with UVES. In the top panel we show the 2PCF of \civ\ and \siiv\ for the lower redshift data \emph{before} merging the lines to mimic \xqr. This is to highlight the small scales that can be probed with the nominal UVES resolution. The bottom panel shows the clustering of \civ\ and \siiv\ based on the \xqr\ data.

The 2PCFs of \civ\ and \siiv\ across both epochs indicate that these metal absorption features are more strongly correlated on small scales compared to a random distribution. This has consistently been found in previous works \citep{Kim2002, Boksenberg2003, Boksenberg2015, Scannapieco2006, Dodorico2010}. Notably, the clustering falls to zero once the separation reaches scales of $300 - 600$~\kms. At larger velocity separations, the clustering of both the \civ\ and \siiv\ absorbers is consistent with what one would expect given a random distribution of metal absorption features. 

As the bottom panel of Figure~\ref{fig:2pcf_xqr30} indicates, the correlation of \civ\ and \siiv\ seem to trace one another well on these scales. This has been found at cosmic noon as inidicated by the top panel of Figure~\ref{fig:2pcf_xqr30} \citep{Boksenberg2003, Scannapieco2006}. For the first time, we highlight that this relationship holds when approaching the end of the epoch of reionisation.

\subsection{Evolution with redshift}
A key motivation to study the clustering of \civ\ and \siiv\ using the E-XQR-30 data is to assess whether there is an appreciable redshift evolution in the 2PCF of these metal ionisation states. This has been tentatively seen when analysing data in the redshift interval $2<z<4$. This has yet to be extended to $z>5$. Thus, we now directly compare the results from E-XQR-30 to those at different epochs. This comparison is made after, when necessary, recalibrating the data from other epochs --- unlike the data shown in the top panel of Figure~\ref{fig:2pcf_xqr30}. 

We initially attempted to assess the potential redshift evolution within the E-XQR-30 survey data. This was approached both by: (1) splitting the sample into two subsamples based on the median absorption redshift, and (2) splitting the sample into two subsamples based on the emission redshift of the quasar. In both cases, we found that they were limited by statistics. There were too few pairs in the subsamples with separations $<1000$~\kms\ to make meaningful comparisons across the redshift intervals. 

\begin{figure}

	\includegraphics[width=\columnwidth]{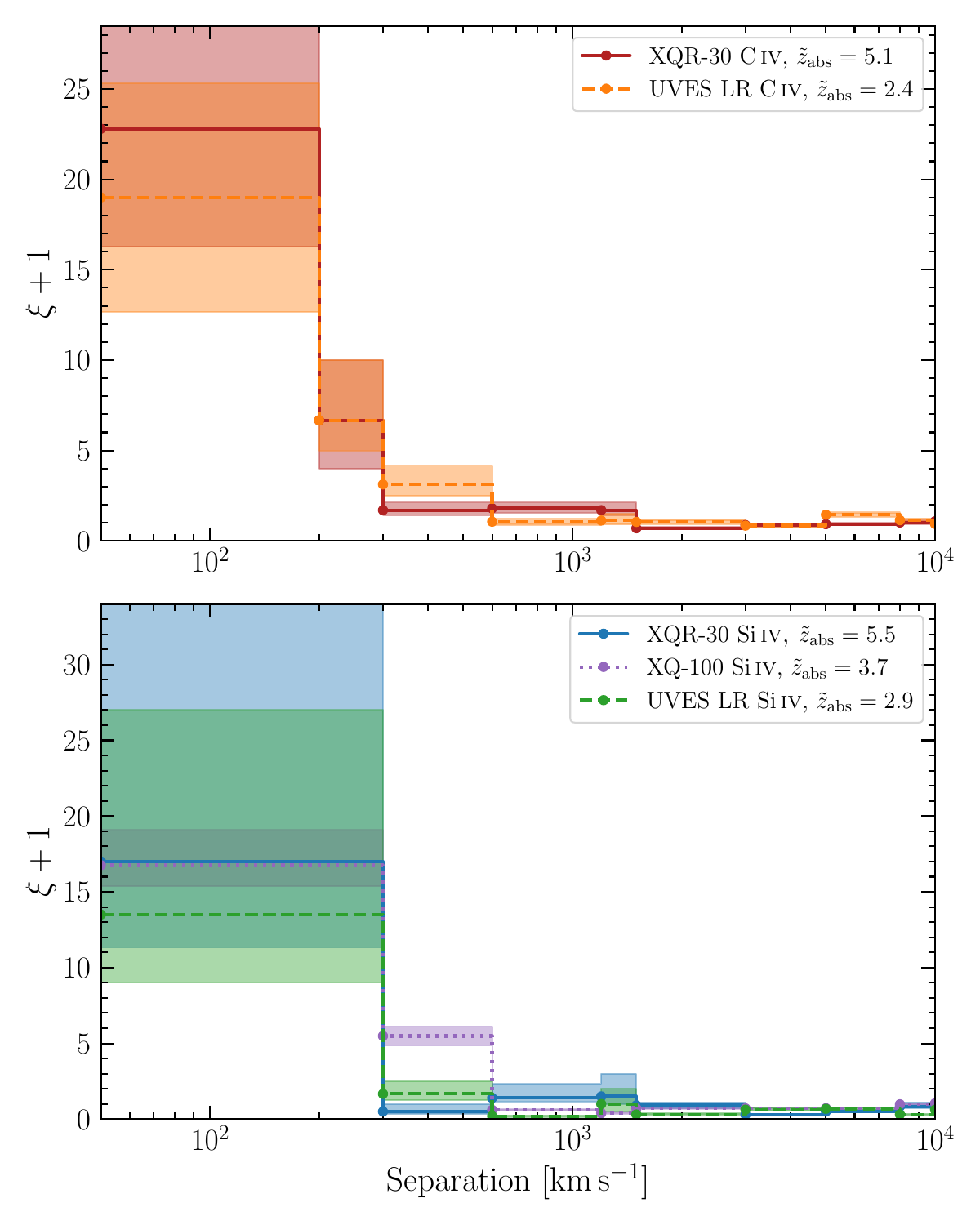}
    \caption{The 2PCF of \civ\ (top) and \siiv\ (bottom) based on the absorbers in the \xqr\ survey now combined with that of the lower redshift `UVES LR' data. The UVES LR data are taken with higher resolution and therefore we have merged the features that would not be resolved with XSHOOTER. We additionally include the 2PCF of \siiv\ from a third dataset, the XQ-100 survey. The shaded region encompasses the interquartile range of the respective distributions. The \siiv\ absorbers in XQ-100 are found at an intermediate epoch between the XQR-30 and UVES LR data. Note that we can probe down to smaller separations (within 200~\kms) with \civ\ due to the increased statistics. }
    \label{fig:2pcf_merged}
\end{figure}

Figure~\ref{fig:2pcf_merged} shows the 2PCF of the \civ\ and \siiv\ absorbers from the E-XQR-30 survey alongside that computed for the other datasets (described in Section~\ref{sec:data}). In this case, the top panel shows the 2PCF of the \civ\ data and the bottom panel shows the 2PCF of the \siiv\ data. Recall that the merged UVES data are denoted as `UVES LR'. When considering \siiv, the probed separations of the first five bins are $\Delta v= [50-300, 300-600, 600-1200, 1200-1500, 1500-3000]$~\kms. When considering \civ\, we include an additional bin between $50-200$~\kms\ to probe the smaller separations more effectively (this adjustment means the second bin spans $200-300$~\kms).

Overall, this figure highlights that there is no significant difference between the 2PCF of either \civ\ or \siiv\ across the epochs traced by the E-XQR-30 ($\tilde{z}_{\rm abs} = 5.1 $) and the UVES LR data ($\tilde{z}_{\rm abs} = 2.4 $). The median values and associated errors are remarkably consistent. Recall that the shaded region indicates the interquartile range of the distribution. We have additionally confirmed that these distributions are consistent with one another given the 1 $\sigma$ confidence intervals. This is true even on the smallest investigated separations: for \civ\ this is within 200~\kms\ while for \siiv\ this is within 300~\kms. This is perhaps surprising given the tentative redshift evolution that has been found in other works. Though, we note that the errors associated with the bins probing the smallest scales are also relatively large. Another possibility to explain the lack of significant evolution is that differences would become increasingly apparent on even smaller scales than those considered in this figure. 

There is a tentative increase in the amplitude of the correlation of the UVES LR data on the scales between $300-600$~\kms compared to that of the E-XQR-30 data that is seen across both \civ\ and \siiv. This may suggest that the \civ\ and \siiv\ systems investigated at $z\sim 2 $ are tracing more massive structures than their counterpart at higher redshift. Note that, at present, the distributions are consistent with one another within the 2$\sigma$ confidence intervals. Perhaps with a larger sample of data, this could be determined more precisely for the smallest scales. Indeed, an increase in the completeness of the sample for lower column density absorbers would be an ideal way to build a larger data sample for this analysis. Given the column density distribution of \civ\ of the \xqr\ data \citep[see Fig. 6 of ][]{Davies2023}, reaching a 50 per cent completeness at $\log_{10}\,N({\rm C \textsc{iv})/ cm^{-2} }=13.00$ would increase the sample size by over one third. Interestingly, the amplitude of the \siiv\ clustering is also distinctly higher in the XQ-100 sample compared to the E-XQR-30 sample when the separation is between $300-600$~\kms. One may have expected the amplitude of this bin to fall between that of E-XQR-30 and the UVES LR sample. It does not and it is consistent with the UVES LR data at the 1$\sigma$ confidence level. 

\begin{figure*}
    \centering
    \includegraphics[width=\linewidth]{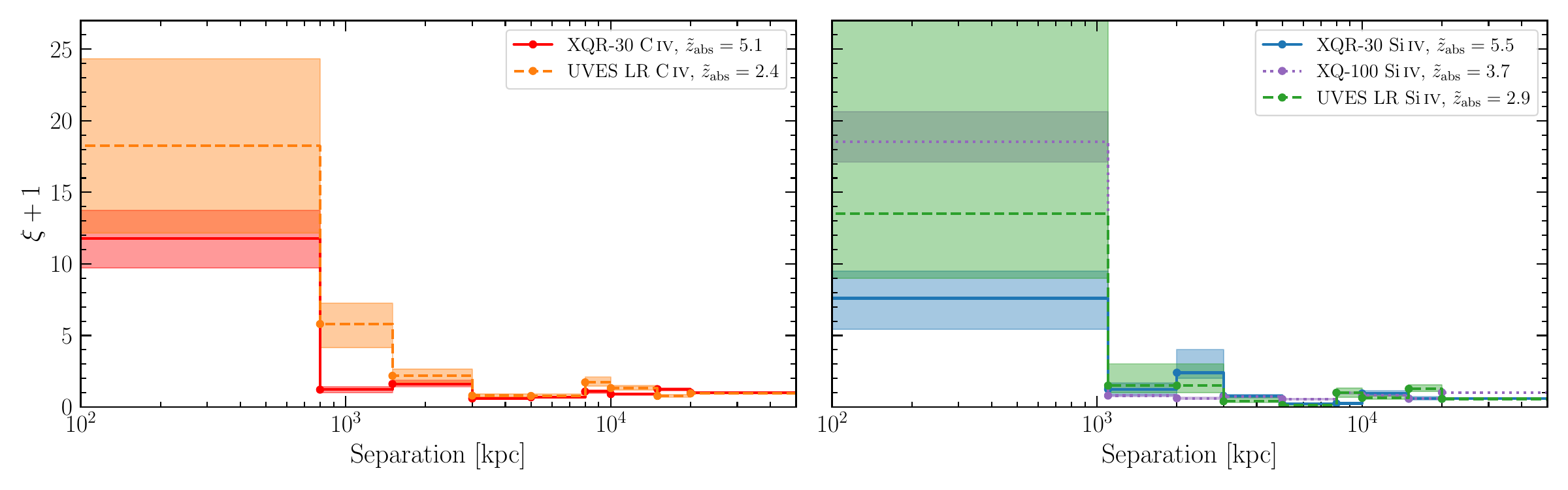}
    \caption{Left: The 2PCF of \civ\ as a function of the separation between pairs in physical kpc space. The legend indicates the different data samples. Right: Same as the left hand side panel but now showing \siiv. For \civ\ the first bin encompasses everything within 800~kpc. For \siiv, the first bin encompasses everything within 1100~kpc. In both panels, the shaded region encompasses the interquartile range of the distributions. } 
    \label{fig:2pcf_kpc}
\end{figure*}

\subsection{Alternative scales}
\label{sec:pkpc}

With these data, we are able to construct a robust estimate of the 2PCF of \civ\ and \siiv\ towards the end of the epoch of reionisation. This is thanks to the extent and quality of the \xqr\ data. However, it also poses an interesting question when it comes to comparing these results to those from other redshift intervals. Traditionally, the 2PCF of absorbers is computed as a function of the separation in velocity space (i.e., \kms). However, the equivalent separation in velocity space will correspond to different physical scales when these epochs are separated by a sufficiently long period in the cosmic history of the Universe. To explore the impact of this, we can assume that the separation between two absorbers in \kms\ is being entirely driven by the motion of the galaxies moving with the Hubble flow. We can then assess what physical scales are being probed by the same velocity separation across different epochs. The minimum resolvable separation in the \xqr\ data is $\Delta v=50$~\kms. Given an average redshift of $z_{\rm abs}=5.1$, this corresponds to physical separation of 90~physical kpc. Conversely, this same separation at the average redshift of the UVES LR data ($z_{\rm abs}=2.4$) is 200~physical kpc. This may explain the lack of evolution when computing the 2PCF in velocity space. 

We now recompute the 2PCF of these metal absorbers in the reference frame of physical distances. Recall, that we assume a Planck cosmology with $H_{0} = 67.4\pm 0.5~{\rm km\,s^{-1}Mpc^{-1}}$, $\Omega_{\rm m} = 0.315 \pm 0.007$ and $h = 0.677$ \citep{Planck2018}. 
Similar to what was found in velocity space, we can investigate down to smaller scales when considering \civ\ alone. However, when considering the behavior of \siiv, we must sample the separations more coarsely.

Figure~\ref{fig:2pcf_kpc} shows the redshift evolution of the 2PCF between $\tilde{z}_{\rm abs} = 5.1 $ and $\tilde{z}_{\rm abs} = 2.4$. This is formatted in the same way as in Figure~\ref{fig:2pcf_merged}. Again, the distributions of plausible $\xi$ values for each sample are formed by generating 1000 realisations of the `fake' absorption lines. The resulting distributions are represented by the median value and the interquartile range. For \civ, we investigate to within 800~kpc in the first bin and 800-1500~kpc in the second bin. For \siiv, the first few bins encompass $50-1100$, $1100- 2000$, and $2000-3000$~kpc. In the left-hand panel, we compare the 2PCF of \civ\ from the E-XQR-30 survey to that of the lower redshift UVES LR data. In this reference frame, there is a tentative difference in the clustering of \civ\ on the smallest scales between these epochs. The amplitude of the UVES LR correlation function is higher than that of the E-XQR-30 data when the separation is within $1500$~kpc. For separations $<800$~kpc, the significance  of the difference is $<1\sigma$. However, there is a $\sim2\sigma$ difference between the amplitude of the clustering across these epochs for separations between $800-1500$~kpc. Most notable, the clustering of the lower redshift absorbers seems to extend to larger separations than that of the \xqr\ data. On these scales, the relative impact of the peculiar velocities of the absorbers should not have a substantial impact.  Indeed, it has been found that (albeit at $z\sim2$) beyond 300 kpc the line-of-sight kinematics of H\,{\sc i} are dominated by Hubble expansion \citep{Chen2020}. Within the scale of the first bin, there is likely to be a contribution from galaxies within a single halo as will be discussed in Section~\ref{sec:sim}.

The right-hand panel of Figure~\ref{fig:2pcf_kpc} shows the 2PCF of \siiv\ based on three data samples that span three redshift intervals. These data show a similar trend to that found for \civ; however, the errors associated with UVES LR data make it challenging to draw firm conclusions on the smallest separations at the lowest redshifts. The \xqr\ and XQ-100 data show a significant difference in the amplitude at the smallest separations ($<1100$~kpc). This follows the trend seen in \civ, highlighting two key results: (1) The \civ\ and \siiv\ trace one another closely when considering separations the reference frame of physical kpc, (2) The amplitude of the clustering of the \civ\ and \siiv\ absorbers increases on the smallest scales as we investigate closer to the present day. This may suggest that the metal absorption line systems are found in relatively denser environments at lower redshift compared to higher redshift. 
Though, we note that his evolution might also be influenced by the evolution of the ionising UV background (UVB). During the epoch of reionisation, the UVB is patchy and dominated by local sources like young galaxies and early quasars. This confines ionised metals to small and localised regions. At cosmic noon, however, the UVB becomes stronger and more uniform, allowing metals to be ionised over larger physical scales, including diffuse regions of the IGM \citep{Khaire2019}. This leads to the enhanced spatial clustering of \civ\ and \siiv\ absorbers as they begin to trace the large-scale structure of the cosmic web. It is important to recognise that the observed evolution reflects the combined impacts of the UVB evolution, large-scale structure growth, and the increasing dispersal of metals into the IGM and CGM over cosmic time.

\section{Discussion}
\label{sec:disc}
Fundamentally, one would like to understand where the gas being analysed arises with respect to the galaxy populations at these different epochs. In principle, the observed evolution in the 2PCF should be reproducible in cosmological hydrodynamical simulations. This would provide an indirect means to associate these metal absorbers with their place within the cosmic web of structure formation. This will be discussed further in Section~\ref{sec:sim}. First we discuss how these results compare to other observations. 

It has been appreciated for multiple decades that the 2PCF of Ly$\alpha$ absorbers shows a significant signal on small velocity scales (i.e., $\Delta v \sim 300$~\kms) with both the amplitude and significance increasing with increasing column density \citep[e.g.][]{Cristiani1997}. At this time, the correlation scale at $z\sim3$ was found to be $150 - 220~h^{-1}$~comoving kpc with an apparent trend of increasing correlation with decreasing redshift. This was reaffirmed with a larger sample of data covering $1.5<z<4$ \citep{Kim2001}. 
\citet{Wolfson2023} have recently extended this investigation to  earlier epochs. They utilise the E-XQR-30 sample to forecast the mean free path of ionising photons at $z> 5.7$ based on the Ly$\alpha$ forest flux autocorrelation function. In \citet{Wolfson2024}, they further compute the autocorrelation function of the Ly$\alpha$ transmission normalised and shifted by the mean transmission in ten redshift bins and find a significant signal for most redshift bins on small scales ($\Delta v\sim200$~\kms). 
We find a significant signal on similar scales when investigating the E-XQR-30 \civ\ absorbers. Thus, these works provide independent yet complementary approaches to studying the clustering of structure using these quasar absorption line data. It is interesting to note that the trend of increasing correlation with decreasing redshift is consistent with our analysis using metal absorbers in place of Ly$\alpha$ absorption.

In addition to Ly$\alpha$ absorption, one may compare how the 2PCF of \civ\ and \siiv\ behaves relative to that of different galaxy types at these epochs. Observationally, the correlation of galaxy types at high redshift has become more accessible thanks to the James Webb Space Telescope (JWST). At more recent epochs, there have been many pioneering studies of galaxy correlations \citep[e.g.,][]{Adelberger2005}. Some of which even compare the galaxy-galaxy and galaxy-absorber correlation (see Section~\ref{sec:gal-abs}).

In particular, JWST-NIRCam has facilitated the photometric selection of Lyman-break galaxies (LBGs) out to unparalleled epochs. \citet{Dalmasso2024} uses the JWST Advanced Deep Extragalactic Survey (JADES) to characterise the galaxy-galaxy correlation function out to $z\sim11$. This work utilises the angular correlation function and the galaxy bias as a measure of the clustering. 
Once complete, the JADES survey will be optimal for identifying robust $z>15$ galaxy candidates \citep{Eisenstein2023}. 
At redshifts similar to that of the E-XQR-30 data, the clustering of [O\,{\sc iii}] and H\,$\beta$ emitting galaxies has been characterised by the Emission-line galaxies and Intergalactic Gas in the Epoch of reionisation (EIGER) survey  \citep{Matthee2023}. Using the angular correlation of the observed galaxies, they find an excess of small scale clustering when the separation is $<2$~arcseconds. These galaxies are found to have stellar masses in the range $M_{\star} = 10^{6.8-10.1}{\rm M_{\odot}}$. While this separation is smaller ($\sim 10$~kpc) than the scales considered in this work, the results are consistent with the trend of increased clustering at smaller separations.

A complementary statistic is the transverse autocorrelation function of absorption features. This provides information beyond the dimensions of an individual line-of-sight. At cosmic noon this has been investigated using \civ\ and Ly$\alpha$ features. For example, at $2<z<3$, the transverse autocorrelation of \civ\ absorbers has been calculated using close quasar pairs \citep{Mintz2022}.  
This work finds that the transverse autocorrelation function appears to level off at $\sim200~h^{-1}$ comoving kpc (see their Fig. 8). Assuming that the \civ\ absorbers identified in the spectra of both members of the quasar pairs arise from the same structure, this indicates that the structures being traced by \civ\ are at least $\sim200~h^{-1}$ comoving kpc in size. A similar study has been conducted using Ly$\alpha$ absorbers detected in the spectra of quasar pairs at $z\sim2$ \citep{Dodorico2006}. This indicates that the clustering of H\,{\sc i} bearing gas is significant up to $\sim 3~h^{-1}$ comoving Mpc. This work also finds good agreement between the line-of-sight and transverse autocorrelation, suggesting that the potential impact of galaxy motions along the line-of-sight are minimal. 
This type of analysis has yet to be extended to the epochs covered by the \xqr\ data.

\subsection{Galaxy-absorber associations}
\label{sec:gal-abs}

To unravel the potential similarities between galaxy-galaxy correlations and absorber-absorber correlations, the detection of gas around galaxies can provide a bridge between these statistics. As with the transverse correlation function, they can also provide a direct determination of the physical extent of gas surrounding galaxies. 
Above $z>4$, MUSE has been utilised to search for galaxies associated with \civ\ absorbers \citep{Diaz2021}. At these redshifts, stronger \civ\ systems are more likely to have LAE galaxies within 200 proper kpc. This association of galaxies and strong \civ\ systems is also seen in the EIGER survey \citep{Kashino2023}. This is again similar to the clustering scales observed through the 2PCF of \civ\ absorbers in the E-XQR-30 sample.

Using JWST, A SPectroscopic survey of biased halos In the reionisation Era (ASPIRE) has found a high incidence rate of galaxies detected within 1000~\kms\ and 350 kpc of known absorbers in the redshift interval $6.0 < z_{\rm abs} < 6.5$. These absorbers are typically detected in both C\,{\sc ii} and \civ\ while the galaxies are detected through [O\,{\sc iii}] and H\,$\beta$ \citep{Zou2024}. Comparing this to the clustering of \civ\ absorbers in the \xqr\ survey, we see an increase in the correlations of absorbers on similar physical kpc scales. There is no evidence to indicate that there is an intrinsic difference between the structures traced by the strong \civ\ systems in the E-XQR-30 data and the galaxy-absorber pairs from ASPIRE. An independent study of galaxies associated with Mg\,{\sc ii} absorbers at this epoch would suggest a highly efficient enrichment of the IGM at the end of reionisation \citep{Bordoloi2024}; this has been found through analysing the same field that has been used in \citet{Kashino2023} while the results presented in \cite{Zou2024} are based on the analysis of four independent fields.

 \subsection{Perspective from galaxy simulations}
 \label{sec:sim}

 Simulations of the formation and evolution of galaxies are an ideal benchmark of comparison for these observational data and the resulting clustering analysis. These comparisons are particularly useful to understand the possible structures giving rise to the observed clustering between absorbers. It is also uniquely suited to disentangling the contribution of different sources to the evolution in the clustering; in simulations, the systematics are well-understood and the contribution of different physical processes (e.g., the evolution of the mean density, the metallicity, and the size of galaxy halos) can be directly determined.

With the latest generation of spectroscopic surveys \citep[e.g., DESI, WEAVE and 4MOST;][]{Dalton2012, deJong2012, DESI2024, Jin2024}, combined with the launch of EUCLID and JWST, there has been an increase in the endeavour to map the correlation function of galaxies at increasingly high redshift through simulations \citep[e.g,][]{Euclid2025a, Euclid2025b, Pizzati2024}. For the purpose of this work, we look to the galaxy clustering as produced through the GAEA semi-analytical galaxy formation model \citep{DeLucia2024}. This model utilises merger trees extracted from the P-Millennium Simulation \citep{Baugh2019} and has recently been used to explore the clustering of galaxies at $0<z<5$ \citep{Fontanot2025}. The model also covers the epochs considered in our analysis, in particular, $z=2.46$ and $z=5.20$. These roughly correspond to the epochs traced by the low redshift UVES LR data and the \xqr\ data. We note that the simulations adopt the same cosmology as used in this work.

 The \textsc{corrfunc} python package can be used to estimate the projected correlation of simulated galaxies based on their position and redshift \citep{Sinha2020}. We consider the correlation of all galaxies in the snapshot as well as the correlation of only the central galaxies. When including satellites, we make no distinction between type-1 or orphan satellites \citep[those whose substructure is stripped below the resolution limit of the simulation;][]{DeLucia2019}. While our analysis relies on the correlation along the line of sight, this can still be used to draw parallels between what we observe and simulations. In this scenario, we are comparing the galaxy-galaxy clustering from simulation snapshots to the absorber-absorber clustering from quasar sightlines.
 
\begin{figure}
    \centering
    \includegraphics[width=\linewidth]{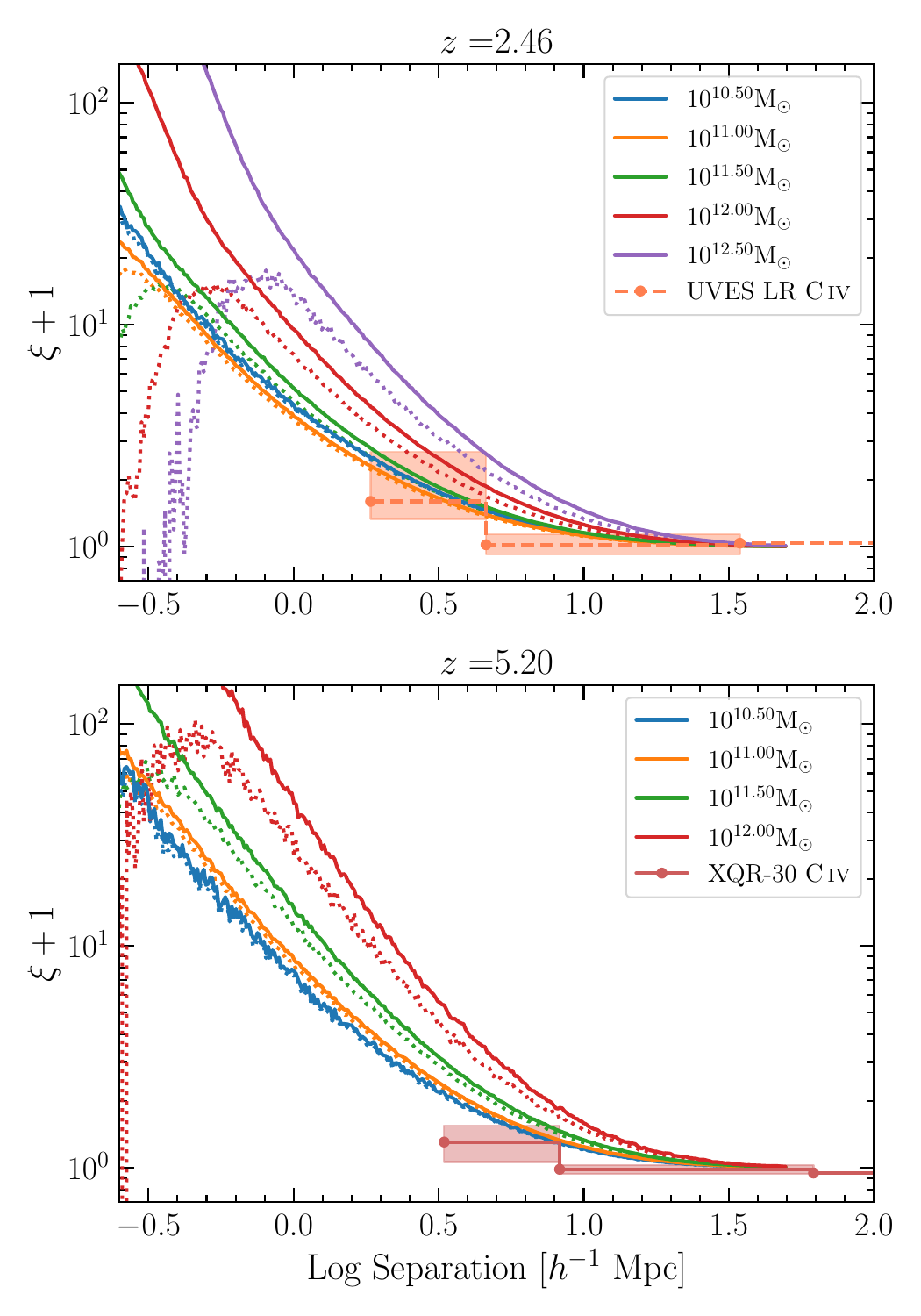}
    \caption{The 3D real space correlation calculated at different epochs using the GAEA simulation suite snapshots. The correlation is calculated using the \textsc{corrfunc} package. The different colors correspond to galaxies with different virial parent halo masses. The top and bottom panels correspond to different epochs as indicated by the titles. The solid curves represent the clustering of all galaxies in the snapshot. The dotted lines represent that of only the central galaxies associated with halos. The observed 2PCF of \civ\ and \siiv\ have been overplotted on these simulation data across different epochs. In this case, we calculate the 2PCF of \civ\ systems (i.e., all the absorption within 200~\kms\ is considered one feature). Again, the shaded regions correspond to the interquartile range of the observed distributions. }
    \label{fig:xi_sim_n_obs}
\end{figure}

Using \textsc{corrfunc}, we compute the 3D real space correlation as a function of the radial position in comoving Mpc. We find that the amplitude of the clustering at a given epoch increases when tracing galaxies with increasing stellar mass. This behaviour is also seen when tracing galaxies embedded in increasingly massive halos as traced by the virial parent halo mass. This can be seen in Figure~\ref{fig:xi_sim_n_obs}. In this figure, we show the correlation of all galaxies associated with a given virial mass (solid line) as well as the correlation of only the central galaxies associated with a given halo (dotted line). As expected, the correlation of the central galaxies drops off at the smallest separations while the correlation between all the detected galaxies does not. This is well-documented in the literature. It can be understood in terms of the halo model --- namely, at small scales the main contribution to the clustering signal is coming from the "one-halo" term, i.e. from galaxies that are sitting within the same halo, while at larger scales what dominates is the "two-halo" term, i.e. the signal is driven by (central) galaxies that are sitting in different haloes. The increase in the amplitude of the correlation is also understood to increase with luminosity (which, for central galaxies, can be a proxy of halo mass). 
For a given virial mass, the amplitude of the correlation is stronger for galaxies at high redshift compared to their lower redshift counterparts. This is true at separation scales larger than $r = 0.3~h^{-1}\rm~Mpc$. We highlight a comparison of each mass bin across the different epochs in Appendix~\ref{appen:xi_v_z}.  

Since the separation between these simulated galaxies is calculated in the reference frame of comoving Mpc, we recompute the 2PCF of \civ\ in this reference frame. The results of this are overplotted on the simulation data in Figure~\ref{fig:xi_sim_n_obs}. In this coordinate system, the \xqr\ data can probe scales as small as $\sim0.36~h^{-1}{\rm~Mpc}$ assuming a median redshift of $z=5.1$. The merged UVES data can probe scales of $\sim0.47~h^{-1}{\rm~Mpc}$. Notably, in this case, we also do not consider the individual absorbers. Instead, we consider the 2PCF of absorption line `systems' as defined in \citet{Davies2023}. These systems are defined as all of the absorption within 200~\kms. We emulate this in the UVES data by merging all of the absorption line features of \civ\ that fall within 200~\kms\ of one another. Following the process described in Section~\ref{sec:data}, these systems are merged into one absorber with a column density equivalent to the sum of the original lines and a redshift that is the column density weighted average of the lines. This choice ensures that the signal we are comparing to the simulations is not dominated by the contribution from absorbers within the same halo. 

The comparison of the observational data and the simulated galaxy correlation in Figure~\ref{fig:xi_sim_n_obs} showcases that the intensity of the correlation as traced by the 3D simulation information of central galaxies is comparable to that traced by \civ\ absorption line systems at $z\sim5.1$ and at $z\sim2.4$; this is true on all the scales probed. The smallest separations that we are testing at each epoch suggest that we are not reaching the scales necessary to probe the correlation of substructures within a given halo. This, in part, motivates our choice to consider absorption line `systems'. The amplitude of the correlation at both $z=2.46$ and $z=5.20$ for the simulated halos with virial masses $M\sim10^{10.5}~{\rm M_{\odot}}$ are the most similar to the trends seen in the real data at these epochs. Overall, we find a consistency between the galaxy-galaxy clustering seen in these simulations and the absorber-absorber clustering seen in the E-XQR-30 data. 

\begin{figure}
    \centering
    \includegraphics[width=\linewidth]{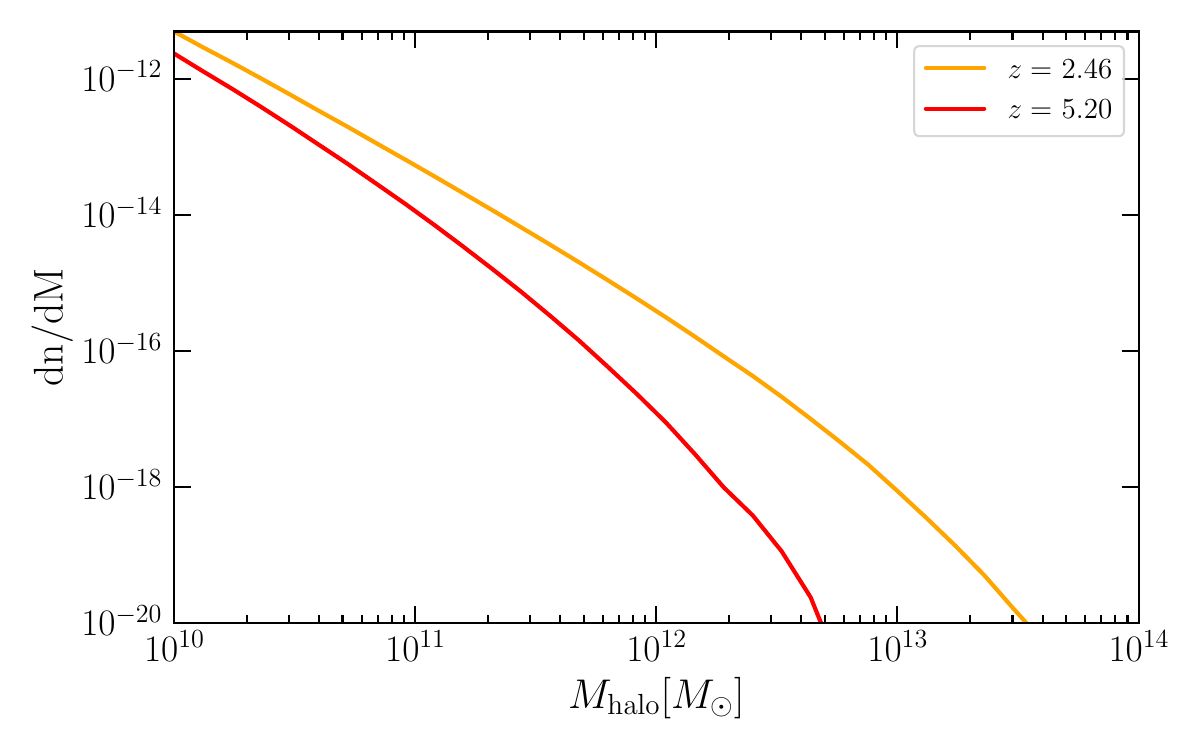}
    \caption{The halo mass function of galaxies from GAEA at the two epochs spanned by the XQR-30 and UVES LR data. This highlights that there are relatively more galaxies of a given halo mass at the epoch explored by the UVES LR data than the epoch explored for the XQR-30 data. This difference becomes more apparent for the highest mass halos.  }
    \label{fig:halom}
\end{figure}

In Figure~\ref{fig:halom} we show the halo mass function of galaxies at the two epochs spanned by the \xqr\ and UVES LR data. Given the density of structures in these simulations, this function is sensitive to halo masses up to $M_{\rm halo} = 10^{14}~\rm M_{\odot}$; this corresponds to a number density per unit mass of $dN/dM\sim 10^{-20}$. As expected, this figure highlights that there are fewer halos with large masses at high redshift compared to lower redshift. It is important to consider how this sensitivity may change across epochs since the structures we are tracing with the \civ\ and \siiv\ absorbers may be associated with the least massive halos.

These qualitative comparisons can provide insight into the nature of the sources that produce the \civ\ systems observed through E-XQR-30.
In the future, the use of synthetic sightlines will allow for a direct comparison with simulations that contain radiative transfer and hydrodynamics \citep{Springel2018, Garaldi2022}. Following the evolution of the 2PCF of \civ\ for different halos in these simulations and comparing that to the real data will allow us to determine the properties of the halos that best reproduce these data. Of particular interest is the typical size of the enriched region around a typical $z = 5.1$ galaxy compared to that at lower redshift. This may produce insight into the size of the CGM across cosmic time. Further, comparisons with the predicted size of pristine/Pop. III enriched bubbles \citep[e.g.,][]{Hartwig2024} could highlight the diversity in the scale of enriched gas at this epoch. This is motivated by the likelihood that the Pop. III enriched regions are likely some of the smaller structures at this epoch compared to those traced by strong \civ\ or \siiv\ absorption.

\section{Conclusions}\label{sec:conc}
We use the E-XQR-30 dataset to investigate the clustering of \civ\ and \siiv\ close to the epoch of reionisation using the 2PCF. We compare these results to that found at later epochs to assess the potential redshift evolution of this clustering. We then draw comparisons with the galaxy clustering properties from the GAEA semi-analytical galaxy formation model. Our main conclusions are as follows:
\begin{enumerate}
    \item We find that the clustering as described by the 2PCF of \civ\ absorbers is indistinguishable between the epoch explored by the E-XQR-30 survey ($\tilde{z}_{\rm abs} = 5.1 $) and a compilation of quasars observed with UVES at cosmic noon ($\tilde{z}_{\rm abs} = 2.4$) when evaluating the separation of absorbers in velocity space.
    \item This is also true when considering the 2PCF of \siiv\ absorbers. For \siiv, we additionally consider a catalogue of absorbers from the XQ-100 survey ($\tilde{z}_{\rm abs} = 3.7$). 
    \item The difference in the 2PCF becomes significant when tracing absorbers in physical kpc scales rather than velocity separation. We find an indication that the amplitude of the correlation increases at lower redshift when considering scales $<1500$~physical kpc. This is consistent with previous works. 
    \item We compare these results to the GAEA semi-analytic galaxy formation model. We compare the evolution of the correlation function from $z=5.20$ to $z=2.46$ as a function of the virial mass of haloes. We find that the correlation signal of the galaxies is broadly consistent with the observed data. This is after accounting for the possibility that the dominant source of the observed clustering between absorbers may be due to the contribution of gas within the same halo. 
    \item We propose that a semi-analytic model of galaxy formation tailored towards understanding the 2PCF of absorbers may provide insight into the size of the CGM and the level of metal enrichment at epochs close to the end of reionisation. 
    \end{enumerate}
The metal absorbers detected towards the end of the reionisation epoch are invaluable tracers of cosmic structure. With the next generation of ground based facilities (e.g., the extremely large telescope; ELT), we will reach an unprecedented sensitivity when observing high redshift quasars. In particular, the ELT-ANDES high resolution spectrograph will be critical to detect the low column density absorbers with a confidence that is inaccessible with current instruments \citep{Dodorico2024}.

\begin{acknowledgements} We thank the anonymous referee for their thoughtful feedback and suggestions that improved the quality of this manuscript. We thank the organisers and participants of the Baryons Beyond Galactic Boundaries 2024 conference for helpful discussions and feedback. LW and VD acknowledges financial support from the Bando Ricerca Fondamentale INAF 2022 Large Grant “XQR-30”.
SEIB is supported by the Deutsche Forschungsgemeinschaft (DFG) under Emmy Noether grant number BO 5771/1-1. 
RLD is supported by the Australian Research Council through the Discovery Early Career Researcher Award (DECRA) Fellowship DE240100136 funded by the Australian Government. 
This research was supported in part by the Australian Research Council Centre of Excellence for All Sky Astrophysics in 3 Dimensions (ASTRO 3D), through project number CE170100013. 

      This paper is based on observations collected at the European Organisation for Astronomical Research in the Southern Hemisphere, Chile. We are grateful to the staff astronomers at the VLT for their assistance with the observations. This research has made use of NASA's Astrophysics Data System.

\end{acknowledgements}

\bibliography{civ_references}

\begin{thebibliography}{72}
\expandafter\ifx\csname natexlab\endcsname\relax\def\natexlab#1{#1}\fi

\bibitem[{{Adelberger} {et~al.}(2005){Adelberger}, {Shapley}, {Steidel}, {Pettini}, {Erb}, \& {Reddy}}]{Adelberger2005}
{Adelberger}, K.~L., {Shapley}, A.~E., {Steidel}, C.~C., {et~al.} 2005, \apj, 629, 636

\bibitem[{{Ba{\~n}ados} {et~al.}(2016){Ba{\~n}ados}, {Venemans}, {Decarli}, {Farina}, {Mazzucchelli}, {Walter}, {Fan}, {Stern}, {Schlafly}, {Chambers}, {Rix}, {Jiang}, {McGreer}, {Simcoe}, {Wang}, {Yang}, {Morganson}, {De Rosa}, {Greiner}, {Balokovi{\'c}}, {Burgett}, {Cooper}, {Draper}, {Flewelling}, {Hodapp}, {Jun}, {Kaiser}, {Kudritzki}, {Magnier}, {Metcalfe}, {Miller}, {Schindler}, {Tonry}, {Wainscoat}, {Waters}, \& {Yang}}]{Banados2016}
{Ba{\~n}ados}, E., {Venemans}, B.~P., {Decarli}, R., {et~al.} 2016, \apjs, 227, 11

\bibitem[{{Baugh} {et~al.}(2019){Baugh}, {Gonzalez-Perez}, {Lagos}, {Lacey}, {Helly}, {Jenkins}, {Frenk}, {Benson}, {Bower}, \& {Cole}}]{Baugh2019}
{Baugh}, C.~M., {Gonzalez-Perez}, V., {Lagos}, C. D.~P., {et~al.} 2019, \mnras, 483, 4922

\bibitem[{{Becker} {et~al.}(2019){Becker}, {Pettini}, {Rafelski}, {D'Odorico}, {Boera}, {Christensen}, {Cupani}, {Ellison}, {Farina}, {Fumagalli}, {L{\'o}pez}, {Neeleman}, {Ryan-Weber}, \& {Worseck}}]{Becker2019}
{Becker}, G.~D., {Pettini}, M., {Rafelski}, M., {et~al.} 2019, \apj, 883, 163

\bibitem[{{Becker} {et~al.}(2009){Becker}, {Rauch}, \& {Sargent}}]{Becker2009}
{Becker}, G.~D., {Rauch}, M., \& {Sargent}, W. L.~W. 2009, \apj, 698, 1010

\bibitem[{{Bird} {et~al.}(2016){Bird}, {Rubin}, {Suresh}, \& {Hernquist}}]{Bird2016}
{Bird}, S., {Rubin}, K. H.~R., {Suresh}, J., \& {Hernquist}, L. 2016, \mnras, 462, 307

\bibitem[{{Bischetti} {et~al.}(2022){Bischetti}, {Feruglio}, {D'Odorico}, {Arav}, {Ba{\~n}ados}, {Becker}, {Bosman}, {Carniani}, {Cristiani}, {Cupani}, {Davies}, {Eilers}, {Farina}, {Ferrara}, {Maiolino}, {Mazzucchelli}, {Mesinger}, {Meyer}, {Onoue}, {Piconcelli}, {Ryan-Weber}, {Schindler}, {Wang}, {Yang}, {Zhu}, \& {Fiore}}]{Bischetti2022}
{Bischetti}, M., {Feruglio}, C., {D'Odorico}, V., {et~al.} 2022, \nat, 605, 244

\bibitem[{{Boksenberg} \& {Sargent}(2015)}]{Boksenberg2015}
{Boksenberg}, A. \& {Sargent}, W. L.~W. 2015, \apjs, 218, 7

\bibitem[{{Boksenberg} {et~al.}(2003){Boksenberg}, {Sargent}, \& {Rauch}}]{Boksenberg2003}
{Boksenberg}, A., {Sargent}, W. L.~W., \& {Rauch}, M. 2003, arXiv e-prints, astro

\bibitem[{{Bordoloi} {et~al.}(2024){Bordoloi}, {Simcoe}, {Matthee}, {Kashino}, {Mackenzie}, {Lilly}, {Eilers}, {Liu}, {DePalma}, {Yue}, \& {P. Naidu}}]{Bordoloi2024}
{Bordoloi}, R., {Simcoe}, R.~A., {Matthee}, J., {et~al.} 2024, \apj, 963, 28

\bibitem[{{Bosman} {et~al.}(2017){Bosman}, {Becker}, {Haehnelt}, {Hewett}, {McMahon}, {Mortlock}, {Simpson}, \& {Venemans}}]{Bosman2017}
{Bosman}, S. E.~I., {Becker}, G.~D., {Haehnelt}, M.~G., {et~al.} 2017, \mnras, 470, 1919

\bibitem[{{Chen} {et~al.}(2020){Chen}, {Steidel}, {Hummels}, {Rudie}, {Dong}, {Trainor}, {Bogosavljevi{\'c}}, {Erb}, {Pettini}, {Reddy}, {Shapley}, {Strom}, {Theios}, {Faucher-Gigu{\`e}re}, {Hopkins}, \& {Kere{\v{s}}}}]{Chen2020}
{Chen}, Y., {Steidel}, C.~C., {Hummels}, C.~B., {et~al.} 2020, \mnras, 499, 1721

\bibitem[{{Christensen} {et~al.}(2023){Christensen}, {Jakobsen}, {Willott}, {Arribas}, {Bunker}, {Charlot}, {Maiolino}, {Marshall}, {Perna}, \& {{\"U}bler}}]{Christensen2023}
{Christensen}, L., {Jakobsen}, P., {Willott}, C., {et~al.} 2023, \aap, 680, A82

\bibitem[{{Cowie} {et~al.}(1995){Cowie}, {Songaila}, {Kim}, \& {Hu}}]{Cowie1995}
{Cowie}, L.~L., {Songaila}, A., {Kim}, T.-S., \& {Hu}, E.~M. 1995, \aj, 109, 1522

\bibitem[{{Cristiani} {et~al.}(1997){Cristiani}, {D'Odorico}, {D'Odorico}, {Fontana}, {Giallongo}, \& {Savaglio}}]{Cristiani1997}
{Cristiani}, S., {D'Odorico}, S., {D'Odorico}, V., {et~al.} 1997, \mnras, 285, 209

\bibitem[{{Dalmasso} {et~al.}(2024){Dalmasso}, {Leethochawalit}, {Trenti}, \& {Boyett}}]{Dalmasso2024}
{Dalmasso}, N., {Leethochawalit}, N., {Trenti}, M., \& {Boyett}, K. 2024, \mnras, 533, 2391

\bibitem[{{Dalton} {et~al.}(2012){Dalton}, {Trager}, {Abrams}, {Carter}, {Bonifacio}, {Aguerri}, {MacIntosh}, {Evans}, {Lewis}, {Navarro}, {Agocs}, {Dee}, {Rousset}, {Tosh}, {Middleton}, {Pragt}, {Terrett}, {Brock}, {Benn}, {Verheijen}, {Cano Infantes}, {Bevil}, {Steele}, {Mottram}, {Bates}, {Gribbin}, {Rey}, {Rodriguez}, {Delgado}, {Guinouard}, {Walton}, {Irwin}, {Jagourel}, {Stuik}, {Gerlofsma}, {Roelfsma}, {Skillen}, {Ridings}, {Balcells}, {Daban}, {Gouvret}, {Venema}, \& {Girard}}]{Dalton2012}
{Dalton}, G., {Trager}, S.~C., {Abrams}, D.~C., {et~al.} 2012, in Society of Photo-Optical Instrumentation Engineers (SPIE) Conference Series, Vol. 8446, Ground-based and Airborne Instrumentation for Astronomy IV, ed. I.~S. {McLean}, S.~K. {Ramsay}, \& H.~{Takami}, 84460P

\bibitem[{{Davies} {et~al.}(2023{\natexlab{a}}){Davies}, {Ryan-Weber}, {D'Odorico}, {Bosman}, {Meyer}, {Becker}, {Cupani}, {Bischetti}, {Sebastian}, {Eilers}, {Farina}, {Wang}, {Yang}, \& {Zhu}}]{Davies2023}
{Davies}, R.~L., {Ryan-Weber}, E., {D'Odorico}, V., {et~al.} 2023{\natexlab{a}}, \mnras, 521, 289

\bibitem[{{Davies} {et~al.}(2023{\natexlab{b}}){Davies}, {Ryan-Weber}, {D'Odorico}, {Bosman}, {Meyer}, {Becker}, {Cupani}, {Keating}, {Bischetti}, {Davies}, {Eilers}, {Farina}, {Haehnelt}, {Pallottini}, \& {Zhu}}]{Davies2023b}
{Davies}, R.~L., {Ryan-Weber}, E., {D'Odorico}, V., {et~al.} 2023{\natexlab{b}}, \mnras, 521, 314

\bibitem[{{de Jong} {et~al.}(2012){de Jong}, {Bellido-Tirado}, {Chiappini}, {Depagne}, {Haynes}, {Johl}, {Schnurr}, {Schwope}, {Walcher}, {Dionies}, {Haynes}, {Kelz}, {Kitaura}, {Lamer}, {Minchev}, {M{\"u}ller}, {Nuza}, {Olaya}, {Piffl}, {Popow}, {Steinmetz}, {Ural}, {Williams}, {Winkler}, {Wisotzki}, {Ansorge}, {Banerji}, {Gonzalez Solares}, {Irwin}, {Kennicutt}, {King}, {McMahon}, {Koposov}, {Parry}, {Sun}, {Walton}, {Finger}, {Iwert}, {Krumpe}, {Lizon}, {Vincenzo}, {Amans}, {Bonifacio}, {Cohen}, {Francois}, {Jagourel}, {Mignot}, {Royer}, {Sartoretti}, {Bender}, {Grupp}, {Hess}, {Lang-Bardl}, {Muschielok}, {B{\"o}hringer}, {Boller}, {Bongiorno}, {Brusa}, {Dwelly}, {Merloni}, {Nandra}, {Salvato}, {Pragt}, {Navarro}, {Gerlofsma}, {Roelfsema}, {Dalton}, {Middleton}, {Tosh}, {Boeche}, {Caffau}, {Christlieb}, {Grebel}, {Hansen}, {Koch}, {Ludwig}, {Quirrenbach}, {Sbordone}, {Seifert}, {Thimm}, {Trifonov}, {Helmi}, {Trager}, {Feltzing}, {Korn}, \& {Boland}}]{deJong2012}
{de Jong}, R.~S., {Bellido-Tirado}, O., {Chiappini}, C., {et~al.} 2012, in Society of Photo-Optical Instrumentation Engineers (SPIE) Conference Series, Vol. 8446, Ground-based and Airborne Instrumentation for Astronomy IV, ed. I.~S. {McLean}, S.~K. {Ramsay}, \& H.~{Takami}, 84460T

\bibitem[{{De Lucia} {et~al.}(2024){De Lucia}, {Fontanot}, {Xie}, \& {Hirschmann}}]{DeLucia2024}
{De Lucia}, G., {Fontanot}, F., {Xie}, L., \& {Hirschmann}, M. 2024, \aap, 687, A68

\bibitem[{{De Lucia} {et~al.}(2019){De Lucia}, {Hirschmann}, \& {Fontanot}}]{DeLucia2019}
{De Lucia}, G., {Hirschmann}, M., \& {Fontanot}, F. 2019, \mnras, 482, 5041

\bibitem[{{Dekker} {et~al.}(2000){Dekker}, {D'Odorico}, {Kaufer}, {Delabre}, \& {Kotzlowski}}]{Dekker2000}
{Dekker}, H., {D'Odorico}, S., {Kaufer}, A., {Delabre}, B., \& {Kotzlowski}, H. 2000, in \procspie, Vol. 4008, Optical and IR Telescope Instrumentation and Detectors, ed. M.~{Iye} \& A.~F. {Moorwood}, 534--545

\bibitem[{{DESI Collaboration} {et~al.}(2024){DESI Collaboration}, {Adame}, {Aguilar}, {Ahlen}, {Alam}, {Aldering}, {Alexander}, {Alfarsy}, {Allende Prieto}, {Alvarez}, {Alves}, {Anand}, {Andrade-Oliveira}, {Armengaud}, {Asorey}, {Avila}, {Aviles}, {Bailey}, {Balaguera-Antol{\'\i}nez}, {Ballester}, {Baltay}, {Bault}, {Bautista}, {Behera}, {Beltran}, {BenZvi}, {Beraldo e Silva}, {Bermejo-Climent}, {Berti}, {Besuner}, {Beutler}, {Bianchi}, {Blake}, {Blum}, {Bolton}, {Brieden}, {Brodzeller}, {Brooks}, {Brown}, {Buckley-Geer}, {Burtin}, {Cabayol-Garcia}, {Cai}, {Canning}, {Cardiel-Sas}, {Carnero Rosell}, {Castander}, {Cervantes-Cota}, {Chabanier}, {Chaussidon}, {Chaves-Montero}, {Chen}, {Chen}, {Chuang}, {Claybaugh}, {Cole}, {Cooper}, {Cuceu}, {Davis}, {Dawson}, {de Belsunce}, {de la Cruz}, {de la Macorra}, {Della Costa}, {de Mattia}, {Demina}, {Demirbozan}, {DeRose}, {Dey}, {Dey}, {Dhungana}, {Ding}, {Ding}, {Doel}, {Doshi}, {Douglass}, {Edge}, {Eftekharzadeh}, {Eisenstein}, {Elliott}, {Ereza}, {Escoffier},
  {Fagrelius}, {Fan}, {Fanning}, {Fawcett}, {Ferraro}, {Flaugher}, {Font-Ribera}, {Forero-Romero}, {Forero-S{\'a}nchez}, {Frenk}, {G{\"a}nsicke}, {Garc{\'\i}a}, {Garc{\'\i}a-Bellido}, {Garcia-Quintero}, {Garrison}, {Gil-Mar{\'\i}n}, {Golden-Marx}, {Gontcho A Gontcho}, {Gonzalez-Morales}, {Gonzalez-Perez}, {Gordon}, {Graur}, {Green}, {Gruen}, {Guy}, {Hadzhiyska}, {Hahn}, {Han}, {Hanif}, {Herrera-Alcantar}, {Honscheid}, {Hou}, {Howlett}, {Huterer}, {Ir{\v{s}}i{\v{c}}}, {Ishak}, {Jacques}, {Jana}, {Jiang}, {Jimenez}, {Jing}, {Joudaki}, {Joyce}, {Jullo}, {Juneau}, {Kara{\c{c}}ayl{\i}}, {Karim}, {Kehoe}, {Kent}, {Khederlarian}, {Kim}, {Kirkby}, {Kisner}, {Kitaura}, {Kizhuprakkat}, {Kneib}, {Koposov}, {Kov{\'a}cs}, {Kremin}, {Krolewski}, {L'Huillier}, {Lahav}, {Lambert}, {Lamman}, {Lan}, {Landriau}, {Lang}, {Lange}, {Lasker}, {Leauthaud}, {Le Guillou}, {Levi}, {Li}, {Linder}, {Lyons}, {Magneville}, {Manera}, {Manser}, {Margala}, {Martini}, {McDonald}, {Medina}, {Medina-Varela}, {Meisner}, {Mena-Fern{\'a}ndez},
  {Meneses-Rizo}, {Mezcua}, {Miquel}, {Montero-Camacho}, {Moon}, {Moore}, {Moustakas}, {Mueller}, {Mundet}, {Mu{\~n}oz-Guti{\'e}rrez}, {Myers}, {Nadathur}, {Napolitano}, {Neveux}, {Newman}, {Nie}, {Nikutta}, {Niz}, {Norberg}, {Noriega}, {Paillas}, {Palanque-Delabrouille}, {Palmese}, {Pan}, {Parkinson}, {Penmetsa}, {Percival}, {P{\'e}rez-Fern{\'a}ndez}, {P{\'e}rez-R{\`a}fols}, {Pieri}, {Poppett}, {Porredon}, {Pothier}, {Prada}, {Pucha}, {Raichoor}, {Ram{\'\i}rez-P{\'e}rez}, {Ramirez-Solano}, {Rashkovetskyi}, {Ravoux}, {Rocher}, {Rockosi}, {Ross}, {Rossi}, {Ruggeri}, {Ruhlmann-Kleider}, {Sabiu}, {Said}, {Saintonge}, {Samushia}, {Sanchez}, {Saulder}, {Schaan}, {Schlafly}, {Schlegel}, {Scholte}, {Schubnell}, {Seo}, {Shafieloo}, {Sharples}, {Sheu}, {Silber}, {Sinigaglia}, {Siudek}, {Slepian}, {Smith}, {Soumagnac}, {Sprayberry}, {Stephey}, {Su{\'a}rez-P{\'e}rez}, {Sun}, {Tan}, {Tarl{\'e}}, {Tojeiro}, {Ure{\~n}a-L{\'o}pez}, {Vaisakh}, {Valcin}, {Valdes}, {Valluri}, {Vargas-Maga{\~n}a}, {Variu}, {Verde}, {Walther},
  {Wang}, {Wang}, {Weaver}, {Weaverdyck}, {Wechsler}, {White}, {Xie}, {Yang}, {Y{\`e}che}, {Yu}, {Yuan}, {Zhang}, {Zhang}, {Zhao}, {Zheng}, {Zhou}, {Zhou}, {Zou}, {Zou}, \& {Zu}}]{DESI2024}
{DESI Collaboration}, {Adame}, A.~G., {Aguilar}, J., {et~al.} 2024, \aj, 168, 58

\bibitem[{{D{\'\i}az} {et~al.}(2011){D{\'\i}az}, {Ryan-Weber}, {Cooke}, {Pettini}, \& {Madau}}]{Diaz2011}
{D{\'\i}az}, C.~G., {Ryan-Weber}, E.~V., {Cooke}, J., {Pettini}, M., \& {Madau}, P. 2011, \mnras, 418, 820

\bibitem[{{D{\'\i}az} {et~al.}(2021){D{\'\i}az}, {Ryan-Weber}, {Karman}, {Caputi}, {Salvadori}, {Crighton}, {Ouchi}, \& {Vanzella}}]{Diaz2021}
{D{\'\i}az}, C.~G., {Ryan-Weber}, E.~V., {Karman}, W., {et~al.} 2021, \mnras, 502, 2645

\bibitem[{{D'Odorico} {et~al.}(2023){D'Odorico}, {Ba{\~n}ados}, {Becker}, {Bischetti}, {Bosman}, {Cupani}, {Davies}, {Farina}, {Ferrara}, {Feruglio}, {Mazzucchelli}, {Ryan-Weber}, {Schindler}, {Sodini}, {Venemans}, {Walter}, {Chen}, {Lai}, {Zhu}, {Bian}, {Campo}, {Carniani}, {Cristiani}, {Davies}, {Decarli}, {Drake}, {Eilers}, {Fan}, {Gaikwad}, {Gallerani}, {Greig}, {Haehnelt}, {Hennawi}, {Keating}, {Kulkarni}, {Mesinger}, {Meyer}, {Neeleman}, {Onoue}, {Pallottini}, {Qin}, {Rojas-Ruiz}, {Satyavolu}, {Sebastian}, {Tripodi}, {Wang}, {Wolfson}, {Yang}, \& {Zanchettin}}]{Dodorico2023}
{D'Odorico}, V., {Ba{\~n}ados}, E., {Becker}, G.~D., {et~al.} 2023, \mnras, 523, 1399

\bibitem[{{D'Odorico} {et~al.}(2024){D'Odorico}, {Bolton}, {Christensen}, {De Cia}, {Zackrisson}, {Kordt}, {Izzo}, {Li}, {Maiolino}, {Marconi}, {Richter}, {Saccardi}, {Salvadori}, {Vanni}, {Feruglio}, {Fumagalli}, {Fynbo}, {Noterdaeme}, {Papaderos}, {P{\'e}roux}, {Verma}, {Di Marcantonio}, {Origlia}, \& {Zanutta}}]{Dodorico2024}
{D'Odorico}, V., {Bolton}, J.~S., {Christensen}, L., {et~al.} 2024, Experimental Astronomy, 58, 21

\bibitem[{{D'Odorico} {et~al.}(2010){D'Odorico}, {Calura}, {Cristiani}, \& {Viel}}]{Dodorico2010}
{D'Odorico}, V., {Calura}, F., {Cristiani}, S., \& {Viel}, M. 2010, \mnras, 401, 2715

\bibitem[{{D'Odorico} {et~al.}(2013){D'Odorico}, {Cupani}, {Cristiani}, {Maiolino}, {Molaro}, {Nonino}, {Centuri{\'o}n}, {Cimatti}, {di Serego Alighieri}, {Fiore}, {Fontana}, {Gallerani}, {Giallongo}, {Mannucci}, {Marconi}, {Pentericci}, {Viel}, \& {Vladilo}}]{Dodorico2013}
{D'Odorico}, V., {Cupani}, G., {Cristiani}, S., {et~al.} 2013, \mnras, 435, 1198

\bibitem[{{D'Odorico} {et~al.}(2022){D'Odorico}, {Finlator}, {Cristiani}, {Cupani}, {Perrotta}, {Calura}, {C{\`e}nturion}, {Becker}, {Berg}, {Lopez}, {Ellison}, \& {Pomante}}]{Dodorico2022}
{D'Odorico}, V., {Finlator}, K., {Cristiani}, S., {et~al.} 2022, \mnras, 512, 2389

\bibitem[{{D'Odorico} {et~al.}(2006){D'Odorico}, {Viel}, {Saitta}, {Cristiani}, {Bianchi}, {Boyle}, {Lopez}, {Maza}, \& {Outram}}]{Dodorico2006}
{D'Odorico}, V., {Viel}, M., {Saitta}, F., {et~al.} 2006, \mnras, 372, 1333

\bibitem[{{Doughty} \& {Finlator}(2023)}]{Doughty2023}
{Doughty}, C.~C. \& {Finlator}, K.~M. 2023, \mnras, 518, 4159

\bibitem[{{Eisenstein} {et~al.}(2023){Eisenstein}, {Johnson}, {Robertson}, {Tacchella}, {Hainline}, {Jakobsen}, {Maiolino}, {Bonaventura}, {Bunker}, {Cameron}, {Cargile}, {Curtis-Lake}, {Hausen}, {Pusk{\'a}s}, {Rieke}, {Sun}, {Willmer}, {Willott}, {Alberts}, {Arribas}, {Baker}, {Baum}, {Bhatawdekar}, {Carniani}, {Charlot}, {Chen}, {Chevallard}, {Curti}, {DeCoursey}, {D'Eugenio}, {de Graaff}, {Egami}, {Helton}, {Ji}, {Jones}, {Kumari}, {L{\"u}tzgendorf}, {Laseter}, {Looser}, {Lyu}, {Maseda}, {Nelson}, {Parlanti}, {Rauscher}, {Rawle}, {Rieke}, {Rix}, {Rujopakarn}, {Sandles}, {Saxena}, {Scholtz}, {Sharpe}, {Shivaei}, {Simmonds}, {Smit}, {Topping}, {{\"U}bler}, {Venturi}, {Williams}, {Witstok}, \& {Woodrum}}]{Eisenstein2023}
{Eisenstein}, D.~J., {Johnson}, B.~D., {Robertson}, B., {et~al.} 2023, arXiv e-prints, arXiv:2310.12340

\bibitem[{{Euclid Collaboration} {et~al.}(2025{\natexlab{a}}){Euclid Collaboration}, {B{\"o}hringer}, {Chon}, {Cucciati}, {Dannerbauer}, {Bolzonella}, {De Lucia}, {Cappi}, {Moscardini}, {Giocoli}, {Castignani}, {Hatch}, {Andreon}, {Ba{\~n}ados}, {Ettori}, {Fontanot}, {Gully}, {Hirschmann}, {Maturi}, {Mei}, {Pozzetti}, {Schlenker}, {Spinelli}, {Aghanim}, {Altieri}, {Auricchio}, {Baccigalupi}, {Baldi}, {Bardelli}, {Bodendorf}, {Bonino}, {Branchini}, {Brescia}, {Brinchmann}, {Camera}, {Capobianco}, {Carbone}, {Carretero}, {Casas}, {Castander}, {Castellano}, {Cavuoti}, {Cimatti}, {Colodro-Conde}, {Congedo}, {Conselice}, {Conversi}, {Copin}, {Courbin}, {Courtois}, {Da Silva}, {Degaudenzi}, {Di Giorgio}, {Dinis}, {Douspis}, {Dubath}, {Duncan}, {Dupac}, {Dusini}, {Farina}, {Farrens}, {Faustini}, {Fosalba}, {Frailis}, {Franceschi}, {Fumana}, {Galeotta}, {Gillis}, {G{\'o}mez-Alvarez}, {Grazian}, {Grupp}, {Haugan}, {Holmes}, {Hormuth}, {Hornstrup}, {Hudelot}, {Jahnke}, {Jhabvala}, {Joachimi}, {Keih{\"a}nen},
  {Kermiche}, {Kiessling}, {Kilbinger}, {Kubik}, {K{\"u}mmel}, {Kunz}, {Kurki-Suonio}, {Ligori}, {Lilje}, {Lindholm}, {Lloro}, {Mainetti}, {Maino}, {Maiorano}, {Mansutti}, {Marggraf}, {Markovic}, {Martinelli}, {Martinet}, {Marulli}, {Massey}, {Maurogordato}, {Medinaceli}, {Mellier}, {Meneghetti}, {Meylan}, {Moresco}, {Niemi}, {Padilla}, {Paltani}, {Pasian}, {Pedersen}, {Pettorino}, {Pires}, {Polenta}, {Poncet}, {Popa}, {Raison}, {Rebolo}, {Renzi}, {Rhodes}, {Riccio}, {Romelli}, {Roncarelli}, {Rossetti}, {Saglia}, {Sakr}, {S{\'a}nchez}, {Sapone}, {Sartoris}, {Schirmer}, {Schneider}, {Scodeggio}, {Secroun}, {Seidel}, {Serrano}, {Sirignano}, {Sirri}, {Stanco}, {Steinwagner}, {Tallada-Cresp{\'\i}}, {Taylor}, {Tereno}, {Toledo-Moreo}, {Torradeflot}, {Tutusaus}, {Vassallo}, {Verdoes Kleijn}, {Veropalumbo}, {Wang}, {Weller}, {Zamorani}, {Zucca}, {Bozzo}, {Burigana}, {Calabrese}, {Di Ferdinando}, {Escartin Vigo}, {Finelli}, {Gracia-Carpio}, {Matthew}, {Mauri}, {P{\"o}ntinen}, {Porciani}, {Scottez}, {Tenti}, {Viel},
  {Wiesmann}, {Akrami}, {Allevato}, {Alvi}, {Anselmi}, {Archidiacono}, {Atrio-Barandela}, {Balaguera-Antolinez}, {Ballardini}, {Blanchard}, {Blot}, {Borgani}, {Bruton}, {Cabanac}, {Calabro}, {Caro}, {Carvalho}, {Castro}, {Chambers}, {Contarini}, {Cooray}, {Costanzi}, {De Caro}, {Desprez}, {D{\'\i}az-S{\'a}nchez}, {Di Domizio}, {Dole}, {Escoffier}, {Ferrari}, {Ferreira}, {Ferrero}, {Fontana}, \& {Fornari}}]{Euclid2025a}
{Euclid Collaboration}, {B{\"o}hringer}, H., {Chon}, G., {et~al.} 2025{\natexlab{a}}, \aap, 693, A59

\bibitem[{{Euclid Collaboration} {et~al.}(2025{\natexlab{b}}){Euclid Collaboration}, {Castander}, {Fosalba}, {Stadel}, {Potter}, {Carretero}, {Tallada-Cresp{\'\i}}, {Pozzetti}, {Bolzonella}, {Mamon}, {Blot}, {Hoffmann}, {Huertas-Company}, {Monaco}, {Gonzalez}, {De Lucia}, {Scarlata}, {Breton}, {Linke}, {Viglione}, {Li}, {Zhai}, {Baghkhani}, {Pardede}, {Neissner}, {Teyssier}, {Crocce}, {Tutusaus}, {Miller}, {Congedo}, {Biviano}, {Hirschmann}, {Pezzotta}, {Aussel}, {Hoekstra}, {Kitching}, {Percival}, {Guzzo}, {Mellier}, {Oesch}, {Bowler}, {Bruton}, {Allevato}, {Gonzalez-Perez}, {Manera}, {Avila}, {Kov{\'a}cs}, {Aghanim}, {Altieri}, {Amara}, {Amendola}, {Andreon}, {Auricchio}, {Baccigalupi}, {Baldi}, {Balestra}, {Bardelli}, {Bender}, {Bernardeau}, {Bodendorf}, {Bonino}, {Branchini}, {Brescia}, {Brinchmann}, {Camera}, {Capobianco}, {Carbone}, {Casas}, {Castellano}, {Castignani}, {Cavuoti}, {Cimatti}, {Colodro-Conde}, {Conselice}, {Conversi}, {Copin}, {Corcione}, {Courbin}, {Courtois}, {Da Silva}, {Degaudenzi},
  {Di Giorgio}, {Dinis}, {Douspis}, {Dubath}, {Duncan}, {Dupac}, {Dusini}, {Ealet}, {Farina}, {Farrens}, {Ferriol}, {Fotopoulou}, {Fourmanoit}, {Frailis}, {Franceschi}, {Franzetti}, {Galeotta}, {Gillard}, {Gillis}, {Giocoli}, {G{\'o}mez-Alvarez}, {Granett}, {Grazian}, {Grupp}, {Haugan}, {Holliman}, {Holmes}, {Hook}, {Hormuth}, {Hornstrup}, {Hudelot}, {Ili{\'c}}, {Jahnke}, {Jhabvala}, {Joachimi}, {Keih{\"a}nen}, {Kermiche}, {Kiessling}, {Kilbinger}, {Kohley}, {Kubik}, {K{\"u}mmel}, {Kunz}, {Kurki-Suonio}, {Lahav}, {Laureijs}, {Le Mignant}, {Liebing}, {Ligori}, {Lilje}, {Lindholm}, {Lloro}, {Maino}, {Maiorano}, {Mansutti}, {Marcin}, {Marggraf}, {Markovic}, {Martinelli}, {Martinet}, {Marulli}, {Massey}, {Masters}, {Maurogordato}, {McCracken}, {Medinaceli}, {Mei}, {Melchior}, {Meneghetti}, {Merlin}, {Meylan}, {Mohr}, {Moresco}, {Moscardini}, {Munari}, {Nakajima}, {Nichol}, {Niemi}, {Padilla}, {Paech}, {Paltani}, {Pasian}, {Peacock}, {Pedersen}, {Pettorino}, {Pires}, {Polenta}, {Poncet}, {Popa}, {Raison},
  {Rebolo}, {Renzi}, {Rhodes}, {Riccio}, {Romelli}, {Roncarelli}, {Rosset}, {Rossetti}, {Rusholme}, {Saglia}, {Sakr}, {S{\'a}nchez}, {Sapone}, {Schewtschenko}, {Schirmer}, {Schneider}, {Schrabback}, {Scodeggio}, {Secroun}, {Sefusatti}, {Seidel}, {Serrano}, {Sirignano}, {Sirri}, {Stanco}, {Starck}, {Steinwagner}, {Taylor}, \& {Teplitz}}]{Euclid2025b}
{Euclid Collaboration}, {Castander}, F.~J., {Fosalba}, P., {et~al.} 2025{\natexlab{b}}, \aap, 697, A5

\bibitem[{{Finlator} {et~al.}(2020){Finlator}, {Doughty}, {Cai}, \& {D{\'\i}az}}]{Finlator2020}
{Finlator}, K., {Doughty}, C., {Cai}, Z., \& {D{\'\i}az}, G. 2020, \mnras, 493, 3223

\bibitem[{{Fiore} {et~al.}(2023){Fiore}, {Ferrara}, {Bischetti}, {Feruglio}, \& {Travascio}}]{Fiore2023}
{Fiore}, F., {Ferrara}, A., {Bischetti}, M., {Feruglio}, C., \& {Travascio}, A. 2023, \apjl, 943, L27

\bibitem[{{Fontanot} {et~al.}(2025){Fontanot}, {De Lucia}, {Xie}, {Hirschmann}, {Baugh}, \& {Helly}}]{Fontanot2025}
{Fontanot}, F., {De Lucia}, G., {Xie}, L., {et~al.} 2025, \aap, 699, A108

\bibitem[{{Garaldi} {et~al.}(2022){Garaldi}, {Kannan}, {Smith}, {Springel}, {Pakmor}, {Vogelsberger}, \& {Hernquist}}]{Garaldi2022}
{Garaldi}, E., {Kannan}, R., {Smith}, A., {et~al.} 2022, \mnras, 512, 4909

\bibitem[{{Hartwig} {et~al.}(2024){Hartwig}, {Lipatova}, {Glover}, \& {Klessen}}]{Hartwig2024}
{Hartwig}, T., {Lipatova}, V., {Glover}, S. C.~O., \& {Klessen}, R.~S. 2024, \mnras, 535, 516

\bibitem[{{Jin} {et~al.}(2024){Jin}, {Trager}, {Dalton}, {Aguerri}, {Drew}, {Falc{\'o}n-Barroso}, {G{\"a}nsicke}, {Hill}, {Iovino}, {Pieri}, {Poggianti}, {Smith}, {Vallenari}, {Abrams}, {Aguado}, {Antoja}, {Arag{\'o}n-Salamanca}, {Ascasibar}, {Babusiaux}, {Balcells}, {Barrena}, {Battaglia}, {Belokurov}, {Bensby}, {Bonifacio}, {Bragaglia}, {Carrasco}, {Carrera}, {Cornwell}, {Dom{\'\i}nguez-Palmero}, {Duncan}, {Famaey}, {Fari{\~n}a}, {Gonzalez}, {Guest}, {Hatch}, {Hess}, {Hoskin}, {Irwin}, {Knapen}, {Koposov}, {Kuchner}, {Laigle}, {Lewis}, {Longhetti}, {Lucatello}, {M{\'e}ndez-Abreu}, {Mercurio}, {Molaeinezhad}, {Mongui{\'o}}, {Morrison}, {Murphy}, {Peralta de Arriba}, {P{\'e}rez}, {P{\'e}rez-R{\`a}fols}, {Pic{\'o}}, {Raddi}, {Romero-G{\'o}mez}, {Royer}, {Siebert}, {Seabroke}, {Som}, {Terrett}, {Thomas}, {Wesson}, {Worley}, {Alfaro}, {Allende Prieto}, {Alonso-Santiago}, {Amos}, {Ashley}, {Balaguer-N{\'u}{\~n}ez}, {Balbinot}, {Bellazzini}, {Benn}, {Berlanas}, {Bernard}, {Best}, {Bettoni}, {Bianco}, {Bishop},
  {Blomqvist}, {Boeche}, {Bolzonella}, {Bonoli}, {Bosma}, {Britavskiy}, {Busarello}, {Caffau}, {Cantat-Gaudin}, {Castro-Ginard}, {Couto}, {Carbajo-Hijarrubia}, {Carter}, {Casamiquela}, {Conrado}, {Corcho-Caballero}, {Costantin}, {Deason}, {de Burgos}, {De Grandi}, {Di Matteo}, {Dom{\'\i}nguez-G{\'o}mez}, {Dorda}, {Drake}, {Dutta}, {Erkal}, {Feltzing}, {Ferr{\'e}-Mateu}, {Feuillet}, {Figueras}, {Fossati}, {Franciosini}, {Frasca}, {Fumagalli}, {Gallazzi}, {Garc{\'\i}a-Benito}, {Gentile Fusillo}, {Gebran}, {Gilbert}, {Gledhill}, {Gonz{\'a}lez Delgado}, {Greimel}, {Guarcello}, {Guerra}, {Gullieuszik}, {Haines}, {Hardcastle}, {Harris}, {Haywood}, {Helmi}, {Hernandez}, {Herrero}, {Hughes}, {Ir{\v{s}}i{\v{c}}}, {Jablonka}, {Jarvis}, {Jordi}, {Kondapally}, {Kordopatis}, {Krogager}, {La Barbera}, {Lam}, {Larsen}, {Lemasle}, {Lewis}, {Lhom{\'e}}, {Lind}, {Lodi}, {Longobardi}, {Lonoce}, {Magrini}, {Ma{\'\i}z Apell{\'a}niz}, {Marchal}, {Marco}, {Martin}, {Matsuno}, {Maurogordato}, {Merluzzi}, {Miralda-Escud{\'e}},
  {Molinari}, {Monari}, {Morelli}, {Mottram}, {Naylor}, {Negueruela}, {O{\~n}orbe}, {Pancino}, {Peirani}, {Peletier}, {Pozzetti}, {Rainer}, {Ramos}, {Read}, {Rossi}, {R{\"o}ttgering}, {Rubi{\~n}o-Mart{\'\i}n}, {Sabater}, {San Juan}, {Sanna}, {Schallig}, {Schiavon}, {Schultheis}, {Serra}, {Shimwell}, {Sim{\'o}n-D{\'\i}az}, {Smith}, {Sordo}, {Sorini}, {Soubiran}, {Starkenburg}, {Steele}, {Stott}, {Stuik}, {Tolstoy}, {Tortora}, {Tsantaki}, {Van der Swaelmen}, {van Weeren}, \& {Vergani}}]{Jin2024}
{Jin}, S., {Trager}, S.~C., {Dalton}, G.~B., {et~al.} 2024, \mnras, 530, 2688

\bibitem[{{Kashino} {et~al.}(2023){Kashino}, {Lilly}, {Matthee}, {Eilers}, {Mackenzie}, {Bordoloi}, \& {Simcoe}}]{Kashino2023}
{Kashino}, D., {Lilly}, S.~J., {Matthee}, J., {et~al.} 2023, \apj, 950, 66

\bibitem[{{Khaire} \& {Srianand}(2019)}]{Khaire2019}
{Khaire}, V. \& {Srianand}, R. 2019, \mnras, 484, 4174

\bibitem[{{Kim} {et~al.}(2001){Kim}, {Cristiani}, \& {D'Odorico}}]{Kim2001}
{Kim}, T.~S., {Cristiani}, S., \& {D'Odorico}, S. 2001, \aap, 373, 757

\bibitem[{{Kim} {et~al.}(2002){Kim}, {Cristiani}, \& {D'Odorico}}]{Kim2002}
{Kim}, T.~S., {Cristiani}, S., \& {D'Odorico}, S. 2002, \aap, 383, 747

\bibitem[{{Lopez} {et~al.}(2024){Lopez}, {Afruni}, {Zamora}, {Tejos}, {Ledoux}, {Hernandez}, {Berg}, {Cortes}, {Urbina}, {Johnston}, {Barrientos}, {Bayliss}, {Cuellar}, {Krogager}, {Noterdaeme}, \& {Solimano}}]{Lopez2024}
{Lopez}, S., {Afruni}, A., {Zamora}, D., {et~al.} 2024, \aap, 691, A356

\bibitem[{{L{\'o}pez} {et~al.}(2016){L{\'o}pez}, {D'Odorico}, {Ellison}, {Becker}, {Christensen}, {Cupani}, {Denney}, {P{\^a}ris}, {Worseck}, {Berg}, {Cristiani}, {Dessauges-Zavadsky}, {Haehnelt}, {Hamann}, {Hennawi}, {Ir{\v{s}}i{\v{c}}}, {Kim}, {L{\'o}pez}, {Lund Saust}, {M{\'e}nard}, {Perrotta}, {Prochaska}, {S{\'a}nchez-Ram{\'\i}rez}, {Vestergaard}, {Viel}, \& {Wisotzki}}]{Lopez2016}
{L{\'o}pez}, S., {D'Odorico}, V., {Ellison}, S.~L., {et~al.} 2016, \aap, 594, A91

\bibitem[{{Lu}(1991)}]{Lu1991}
{Lu}, L. 1991, \apj, 379, 99

\bibitem[{{Matthee} {et~al.}(2023){Matthee}, {Mackenzie}, {Simcoe}, {Kashino}, {Lilly}, {Bordoloi}, \& {Eilers}}]{Matthee2023}
{Matthee}, J., {Mackenzie}, R., {Simcoe}, R.~A., {et~al.} 2023, \apj, 950, 67

\bibitem[{{Mintz} {et~al.}(2022){Mintz}, {Rafelski}, {Jorgenson}, {Fumagalli}, {Dutta}, {Martin}, {Lusso}, {Rubin}, \& {O'Meara}}]{Mintz2022}
{Mintz}, A., {Rafelski}, M., {Jorgenson}, R.~A., {et~al.} 2022, \aj, 164, 51

\bibitem[{{P{\'e}roux} \& {Howk}(2020)}]{Peroux2020}
{P{\'e}roux}, C. \& {Howk}, J.~C. 2020, \araa, 58, 363

\bibitem[{{Pettini} {et~al.}(2003){Pettini}, {Madau}, {Bolte}, {Prochaska}, {Ellison}, \& {Fan}}]{Pettini2003}
{Pettini}, M., {Madau}, P., {Bolte}, M., {et~al.} 2003, \apj, 594, 695

\bibitem[{{Pizzati} {et~al.}(2024){Pizzati}, {Hennawi}, {Schaye}, {Schaller}, {Eilers}, {Wang}, {Frenk}, {Elbers}, {Helly}, {Mackenzie}, {Matthee}, {Bordoloi}, {Kashino}, {Naidu}, \& {Yue}}]{Pizzati2024}
{Pizzati}, E., {Hennawi}, J.~F., {Schaye}, J., {et~al.} 2024, \mnras, 534, 3155

\bibitem[{{Planck Collaboration} {et~al.}(2020){Planck Collaboration}, {Aghanim}, {Akrami}, {Ashdown}, {Aumont}, {Baccigalupi}, {Ballardini}, {Banday}, {Barreiro}, {Bartolo}, {Basak}, {Battye}, {Benabed}, {Bernard}, {Bersanelli}, {Bielewicz}, {Bock}, {Bond}, {Borrill}, {Bouchet}, {Boulanger}, {Bucher}, {Burigana}, {Butler}, {Calabrese}, {Cardoso}, {Carron}, {Challinor}, {Chiang}, {Chluba}, {Colombo}, {Combet}, {Contreras}, {Crill}, {Cuttaia}, {de Bernardis}, {de Zotti}, {Delabrouille}, {Delouis}, {Di Valentino}, {Diego}, {Dor{\'e}}, {Douspis}, {Ducout}, {Dupac}, {Dusini}, {Efstathiou}, {Elsner}, {En{\ss}lin}, {Eriksen}, {Fantaye}, {Farhang}, {Fergusson}, {Fernandez-Cobos}, {Finelli}, {Forastieri}, {Frailis}, {Fraisse}, {Franceschi}, {Frolov}, {Galeotta}, {Galli}, {Ganga}, {G{\'e}nova-Santos}, {Gerbino}, {Ghosh}, {Gonz{\'a}lez-Nuevo}, {G{\'o}rski}, {Gratton}, {Gruppuso}, {Gudmundsson}, {Hamann}, {Handley}, {Hansen}, {Herranz}, {Hildebrandt}, {Hivon}, {Huang}, {Jaffe}, {Jones}, {Karakci}, {Keih{\"a}nen},
  {Keskitalo}, {Kiiveri}, {Kim}, {Kisner}, {Knox}, {Krachmalnicoff}, {Kunz}, {Kurki-Suonio}, {Lagache}, {Lamarre}, {Lasenby}, {Lattanzi}, {Lawrence}, {Le Jeune}, {Lemos}, {Lesgourgues}, {Levrier}, {Lewis}, {Liguori}, {Lilje}, {Lilley}, {Lindholm}, {L{\'o}pez-Caniego}, {Lubin}, {Ma}, {Mac{\'\i}as-P{\'e}rez}, {Maggio}, {Maino}, {Mandolesi}, {Mangilli}, {Marcos-Caballero}, {Maris}, {Martin}, {Martinelli}, {Mart{\'\i}nez-Gonz{\'a}lez}, {Matarrese}, {Mauri}, {McEwen}, {Meinhold}, {Melchiorri}, {Mennella}, {Migliaccio}, {Millea}, {Mitra}, {Miville-Desch{\^e}nes}, {Molinari}, {Montier}, {Morgante}, {Moss}, {Natoli}, {N{\o}rgaard-Nielsen}, {Pagano}, {Paoletti}, {Partridge}, {Patanchon}, {Peiris}, {Perrotta}, {Pettorino}, {Piacentini}, {Polastri}, {Polenta}, {Puget}, {Rachen}, {Reinecke}, {Remazeilles}, {Renzi}, {Rocha}, {Rosset}, {Roudier}, {Rubi{\~n}o-Mart{\'\i}n}, {Ruiz-Granados}, {Salvati}, {Sandri}, {Savelainen}, {Scott}, {Shellard}, {Sirignano}, {Sirri}, {Spencer}, {Sunyaev}, {Suur-Uski}, {Tauber}, {Tavagnacco},
  {Tenti}, {Toffolatti}, {Tomasi}, {Trombetti}, {Valenziano}, {Valiviita}, {Van Tent}, {Vibert}, {Vielva}, {Villa}, {Vittorio}, {Wandelt}, {Wehus}, {White}, {White}, {Zacchei}, \& {Zonca}}]{Planck2018}
{Planck Collaboration}, {Aghanim}, N., {Akrami}, Y., {et~al.} 2020, \aap, 641, A6

\bibitem[{{Ryan-Weber} {et~al.}(2006){Ryan-Weber}, {Pettini}, \& {Madau}}]{RyanWeber2006}
{Ryan-Weber}, E.~V., {Pettini}, M., \& {Madau}, P. 2006, \mnras, 371, L78

\bibitem[{{Ryan-Weber} {et~al.}(2009){Ryan-Weber}, {Pettini}, {Madau}, \& {Zych}}]{RyanWeber2009}
{Ryan-Weber}, E.~V., {Pettini}, M., {Madau}, P., \& {Zych}, B.~J. 2009, \mnras, 395, 1476

\bibitem[{{Scannapieco} {et~al.}(2006){Scannapieco}, {Pichon}, {Aracil}, {Petitjean}, {Thacker}, {Pogosyan}, {Bergeron}, \& {Couchman}}]{Scannapieco2006}
{Scannapieco}, E., {Pichon}, C., {Aracil}, B., {et~al.} 2006, \mnras, 365, 615

\bibitem[{{Sebastian} {et~al.}(2024){Sebastian}, {Ryan-Weber}, {Davies}, {Becker}, {Keating}, {D'Odorico}, {Meyer}, {Bosman}, {Cupani}, {Kulkarni}, {Haehnelt}, {Lai}, {Eilers}, {Bischetti}, \& {Gallerani}}]{Sebastian2024}
{Sebastian}, A.~M., {Ryan-Weber}, E., {Davies}, R.~L., {et~al.} 2024, \mnras, 530, 1829

\bibitem[{{Simcoe}(2006)}]{Simcoe2006}
{Simcoe}, R.~A. 2006, \apj, 653, 977

\bibitem[{{Sinha} \& {Garrison}(2020)}]{Sinha2020}
{Sinha}, M. \& {Garrison}, L.~H. 2020, \mnras, 491, 3022

\bibitem[{{Songaila}(2001)}]{Songaila2001}
{Songaila}, A. 2001, \apjl, 561, L153

\bibitem[{{Springel} {et~al.}(2018){Springel}, {Pakmor}, {Pillepich}, {Weinberger}, {Nelson}, {Hernquist}, {Vogelsberger}, {Genel}, {Torrey}, {Marinacci}, \& {Naiman}}]{Springel2018}
{Springel}, V., {Pakmor}, R., {Pillepich}, A., {et~al.} 2018, \mnras, 475, 676

\bibitem[{{Steidel} {et~al.}(2010){Steidel}, {Erb}, {Shapley}, {Pettini}, {Reddy}, {Bogosavljevi{\'c}}, {Rudie}, \& {Rakic}}]{Steidel2010}
{Steidel}, C.~C., {Erb}, D.~K., {Shapley}, A.~E., {et~al.} 2010, \apj, 717, 289

\bibitem[{{Tescari} {et~al.}(2011){Tescari}, {Viel}, {D'Odorico}, {Cristiani}, {Calura}, {Borgani}, \& {Tornatore}}]{Tescari2011}
{Tescari}, E., {Viel}, M., {D'Odorico}, V., {et~al.} 2011, \mnras, 411, 826

\bibitem[{{Tortosa} {et~al.}(2024){Tortosa}, {Zappacosta}, {Piconcelli}, {Bischetti}, {Done}, {Miniutti}, {Saccheo}, {Vietri}, {Bongiorno}, {Brusa}, {Carniani}, {Chilingarian}, {Civano}, {Cristiani}, {D'Odorico}, {Elvis}, {Fan}, {Feruglio}, {Fiore}, {Gallerani}, {Giallongo}, {Gilli}, {Grazian}, {Guainazzi}, {Haardt}, {Luminari}, {Maiolino}, {Menci}, {Nicastro}, {Petrucci}, {Puccetti}, {Salvestrini}, {Schneider}, {Testa}, {Tombesi}, {Tripodi}, {Valiante}, {Vallini}, {Vanzella}, {Vasylenko}, {Vignali}, {Vito}, {Volonteri}, \& {La Franca}}]{Tortosa2024}
{Tortosa}, A., {Zappacosta}, L., {Piconcelli}, E., {et~al.} 2024, \aap, 691, A235

\bibitem[{{Tumlinson} {et~al.}(2017){Tumlinson}, {Peeples}, \& {Werk}}]{Tumlinson2017}
{Tumlinson}, J., {Peeples}, M.~S., \& {Werk}, J.~K. 2017, \araa, 55, 389

\bibitem[{{Vernet} {et~al.}(2011){Vernet}, {Dekker}, {D'Odorico}, {Kaper}, {Kjaergaard}, {Hammer}, {Randich}, {Zerbi}, {Groot}, {Hjorth}, {Guinouard}, {Navarro}, {Adolfse}, {Albers}, {Amans}, {Andersen}, {Andersen}, {Binetruy}, {Bristow}, {Castillo}, {Chemla}, {Christensen}, {Conconi}, {Conzelmann}, {Dam}, {de Caprio}, {de Ugarte Postigo}, {Delabre}, {di Marcantonio}, {Downing}, {Elswijk}, {Finger}, {Fischer}, {Flores}, {Fran{\c{c}}ois}, {Goldoni}, {Guglielmi}, {Haigron}, {Hanenburg}, {Hendriks}, {Horrobin}, {Horville}, {Jessen}, {Kerber}, {Kern}, {Kiekebusch}, {Kleszcz}, {Klougart}, {Kragt}, {Larsen}, {Lizon}, {Lucuix}, {Mainieri}, {Manuputy}, {Martayan}, {Mason}, {Mazzoleni}, {Michaelsen}, {Modigliani}, {Moehler}, {M{\o}ller}, {Norup S{\o}rensen}, {N{\o}rregaard}, {P{\'e}roux}, {Patat}, {Pena}, {Pragt}, {Reinero}, {Rigal}, {Riva}, {Roelfsema}, {Royer}, {Sacco}, {Santin}, {Schoenmaker}, {Spano}, {Sweers}, {Ter Horst}, {Tintori}, {Tromp}, {van Dael}, {van der Vliet}, {Venema}, {Vidali}, {Vinther}, {Vola},
  {Winters}, {Wistisen}, {Wulterkens}, \& {Zacchei}}]{Vernet2011}
{Vernet}, J., {Dekker}, H., {D'Odorico}, S., {et~al.} 2011, \aap, 536, A105

\bibitem[{{Wolfson} {et~al.}(2024){Wolfson}, {Hennawi}, {Bosman}, {Davies}, {Luki{\'c}}, {Becker}, {Chen}, {Cupani}, {D'Odorico}, {Eilers}, {Haehnelt}, {Keating}, {Kulkarni}, {Lai}, {Mesinger}, {Walter}, \& {Zhu}}]{Wolfson2024}
{Wolfson}, M., {Hennawi}, J.~F., {Bosman}, S. E.~I., {et~al.} 2024, \mnras, 531, 3069

\bibitem[{{Wolfson} {et~al.}(2023){Wolfson}, {Hennawi}, {Davies}, \& {O{\~n}orbe}}]{Wolfson2023}
{Wolfson}, M., {Hennawi}, J.~F., {Davies}, F.~B., \& {O{\~n}orbe}, J. 2023, \mnras, 521, 4056

\bibitem[{{Yang} {et~al.}(2023){Yang}, {Fan}, {Gupta}, {Myers}, {Palanque-Delabrouille}, {Wang}, {Y{\`e}che}, {Aguilar}, {Ahlen}, {Alexander}, {Brooks}, {Dawson}, {de la Macorra}, {Dey}, {Dhungana}, {Fanning}, {Font-Ribera}, {Gontcho}, {Guy}, {Honscheid}, {Juneau}, {Kisner}, {Kremin}, {Le Guillou}, {Levi}, {Magneville}, {Martini}, {Meisner}, {Miquel}, {Moustakas}, {Nie}, {Percival}, {Poppett}, {Prada}, {Schlafly}, {Tarl{\'e}}, {Vargas Magana}, {Weaver}, {Wechsler}, {Zhou}, {Zhou}, \& {Zou}}]{Yang2023}
{Yang}, J., {Fan}, X., {Gupta}, A., {et~al.} 2023, \apjs, 269, 27

\bibitem[{{Zou} {et~al.}(2024){Zou}, {Cai}, {Wang}, {Fan}, {Champagne}, {Hennawi}, {Schindler}, {Farina}, {Yang}, {Inayoshi}, {Ba{\~n}ados}, {Bosman}, {Li}, {Lin}, {Wu}, {Sun}, {Guo}, {Kulkuarni}, {Habouzit}, {Charlot}, {Chevallard}, {Connor}, {Eilers}, {Jiang}, {Jin}, {Kakiichi}, {Li}, {Meyer}, {Walter}, \& {Zhang}}]{Zou2024}
{Zou}, S., {Cai}, Z., {Wang}, F., {et~al.} 2024, \apjl, 963, L28

\end{thebibliography}
\bibliographystyle{aa}

\begin{appendix}
    
\section{Clustering over time for a given halo mass}
\label{appen:xi_v_z}

\begin{figure}
    \centering
    \includegraphics[width=\linewidth]{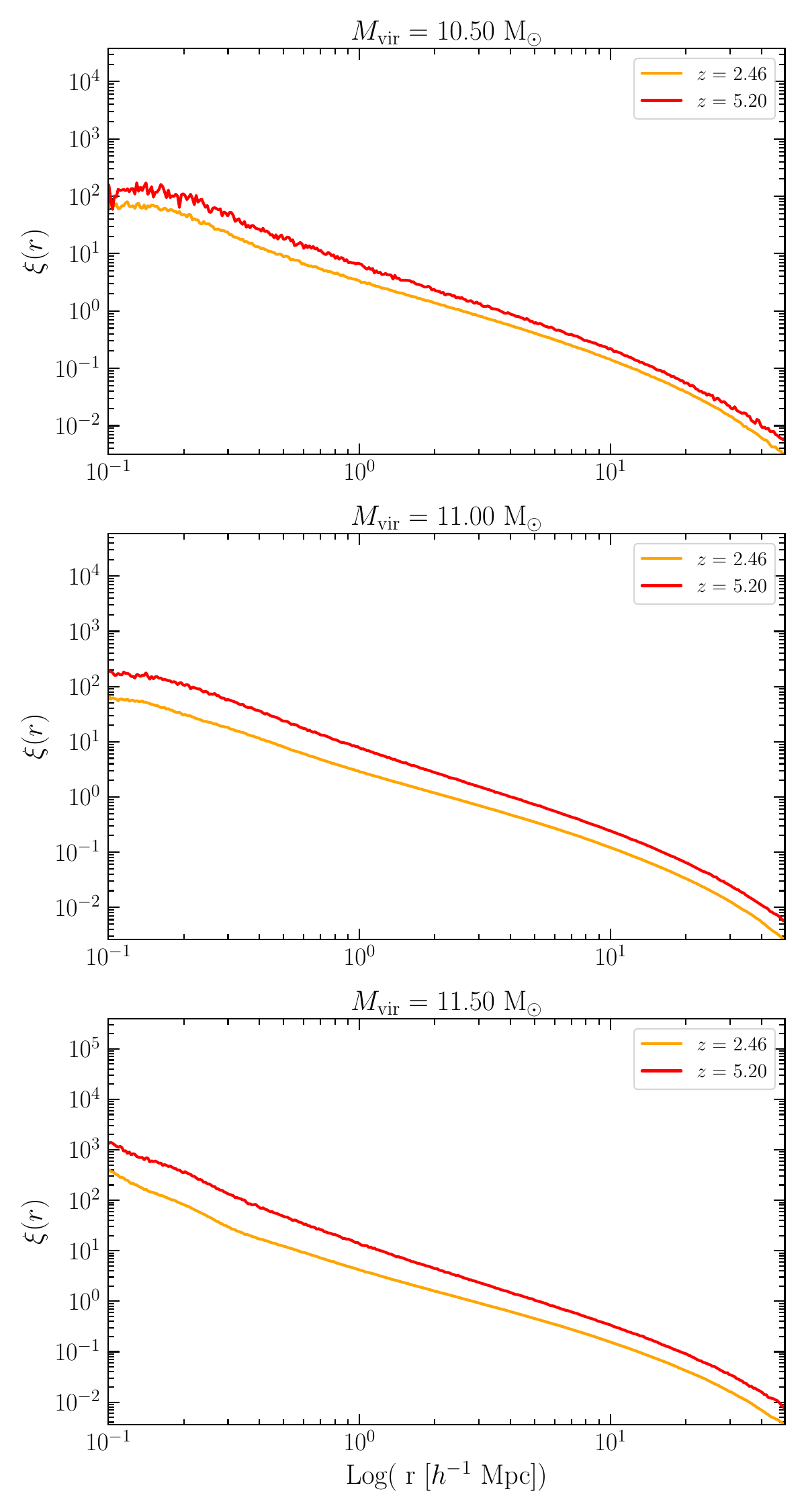}
    \caption{The clustering of structures across different cosmic epochs. The different rows show this clustering for galaxies within different halo mass bins.}
    \label{fig:xi_v_z}
\end{figure}

In this appendix we highlight how the clustering of structure evolves from the epoch of reionisation to cosmic noon for galaxies of increasing virial parent halo masses. This is highlighted in Figure~\ref{fig:xi_v_z}. The increasing amplitude of the correlation at relatively higher redshift is well documented in the literature. 

\end{appendix}

\end{document}